\documentclass[final,times,twocolumn]{elsarticle}
\usepackage{epsfig}
\usepackage{amsmath,amssymb,bm}
\usepackage{graphics}
\usepackage{epsfig}
\usepackage{graphicx}
\usepackage{verbatim}

\usepackage{widetext}

\usepackage{mathtools}
\usepackage[dvipsnames]{xcolor}

\journal{Chaos, Solitons \& Fractals}

\def\ER{Erd\H{o}s-R\'enyi }

\begin{document}

\begin{frontmatter}

\title{Weak percolation on multiplex networks with overlapping edges}
\author{G.~J.~Baxter}
\author{R.~A.~da~Costa}
\author{S.~N. Dorogovtsev}
\author{J.~F.~F. Mendes}
\address{{Department of Physics \& I3N, University of Aveiro},
            {Campus Universit\'ario de Santiago}, 
            {Aveiro},
            {3810-193}, 
           {Portugal}}
            

\begin{abstract}
We solve the weak percolation problem for multiplex networks with overlapping edges. 
In weak percolation, a vertex belongs to a connected component if at least one of its 
neighbors in each of the layers is in this component. This is a weaker condition than for a mutually
connected component in interdependent networks, in which any two vertices must be connected by a
path within each of the layers. 
The effect of the overlaps on weak percolation turns out to be opposite to that on the giant mutually connected component.
While for the giant mutually connected component, overlaps do not change the critical phenomena,  our theory shows that in two layers any (nonzero) concentration of overlaps drives the weak percolation transition to the ordinary percolation universality class.
In three layers, the phase diagram of the problem contains two lines---of a continuous phase transition and of a discontinuous one---connected in various ways depending on how the layers overlap. In the case of only doubled overlapped edges, two of the end points of these lines coincide, resulting in  a tricritical point like that seen in heterogeneous $k$-core percolation.
\end{abstract}







\end{frontmatter}


\section{Introduction}
\label{s1} 

Networks are a convenient representation of the heterogeneous interactions present in many complex systems, and their percolation properties, in turn, give insight into their structure and the dynamics of processes occurring on them \cite{newman2010networks,dorogovtsev2008critical, dorogovtsev2022the}.
The generalisation of percolation to multi-layer networks allows for the representation of systems consisting of multiple interdependent sub-systems, or multiple types of interactions. 
Such interdependent or multiplex networks has revealed has generated significant attention in recent years, revealing a range of novel critical phenomena \cite{buldyrev2010catastrophic, baxter2012avalanche, baxter2014weak, bianconi2018multilayer, baxter2020exotic}. 

In fact the generalisation of percolation to such networks is not unique. There are two basic conceptions of (giant) components in networks with a multiplex architecture and networks of networks:
\\ 
(i) In the standard interdependent network definition, one looks for a giant mutually connected component (giant viable cluster) \cite{buldyrev2010catastrophic}. Such a cluster is defined by a global (non-local) condition: Each pair of vertices in the mutual component must be connected by at least one path of each existing color. 
To identify mutually connected components, a nonlocal pruning process must be used. For example, 
find the giant connected components in each of the layers and remove everything apart from their overlap. These removals may alter the respective giant components, so the process must be iterated until a stable solution is found.

(ii) Weak percolation. This definition involves a local condition: Each vertex must maintain a connection to at least one other included vertex in each layer \cite{baxter2014weak, baxter2021weak}. That is, the condition that there must exist a path in each layer to all other vertices in the cluster is relaxed. 
 Weak percolation clusters may be identified by a local pruning process, progressively removing all vertices not satisfying the condition. This process is therefore much more direct, and is similar to the single-layer $k$-core pruning process \cite{baxter2015critical}. In fact, weak multiplex percolation is equivalent 
 to the particular case of {\bf k}-core percolation (the generalization of $k$-core to multiple layers) with the vector $\text{\bf k} = (1,1,\ldots,1)$  \cite{azimi2014k}, see also Ref. \cite{di2017cascading}.

Each of these formulations is appropriate for different applications. The interdependent networks definition, problem (i), models situations in which the diversity of connectivity is important. Weak percolation, problem (ii), on the other hand, models situations in which the diversity of local environment matters, that is, for survival, you must have diverse neighbors. 

Problem (i) leads to a discontinuous hybrid phase transition in two or more layers.
Problem (ii) for a two-layer network leads to a continuous phase transition with exponent $2$ if degree distributions decay sufficiently rapidly \cite{baxter2020exotic}, and the transition becomes a discontinuous hybrid one in three or more layers \cite{baxter2021weak}.

Most treatments of these problems have considered large sparse random networks, in which the probability that a pair of vertices is linked in more than one layer vanishes in the infinite size limit \cite{buldyrev2010catastrophic, shao2011cascade, baxter2012avalanche, son2012percolation}. 
However in real systems, constraints such as the spatial location of vertices means that there is a significant probability to find such co-located or overlapping edges.
The effect of such overlaps on the giant mutually connected component in interdependent networks attracted particular attention. It was shown that overlaps cannot destroy the discontinuous nature of the transition \cite{hu2013percolation, min2015link,baxter2016correlated, cellai2016message}.
Only a 100\% concentration of overlaps can produce the normal percolation phase transition. 
This provides a simple phase diagram where two phases with and without the giant mutually connected component are separated by the hybrid transition line. 
t

\begin{figure}
\begin{center}
\includegraphics[width=0.8\columnwidth]{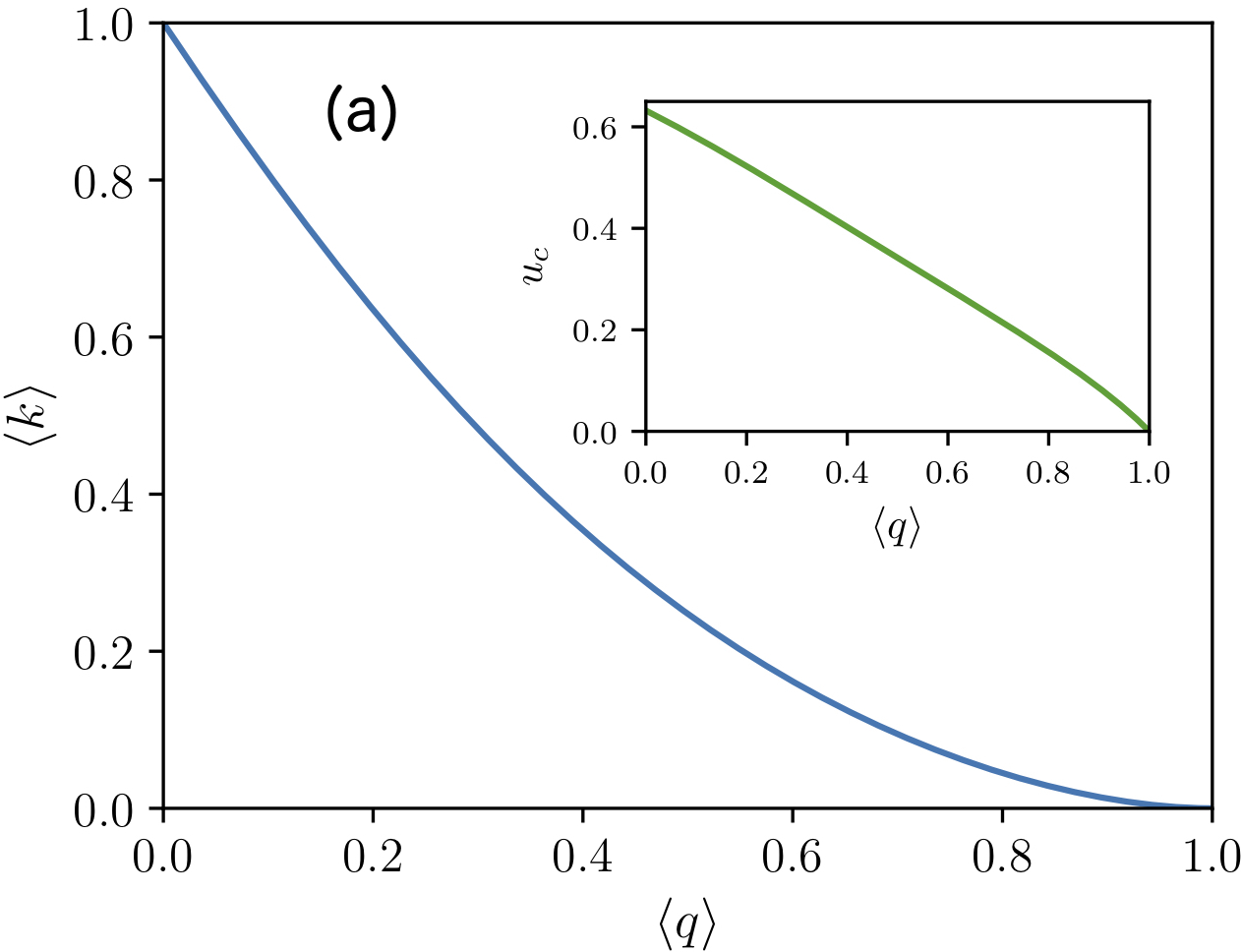} \\[2pt]
\includegraphics[width=0.8\columnwidth]{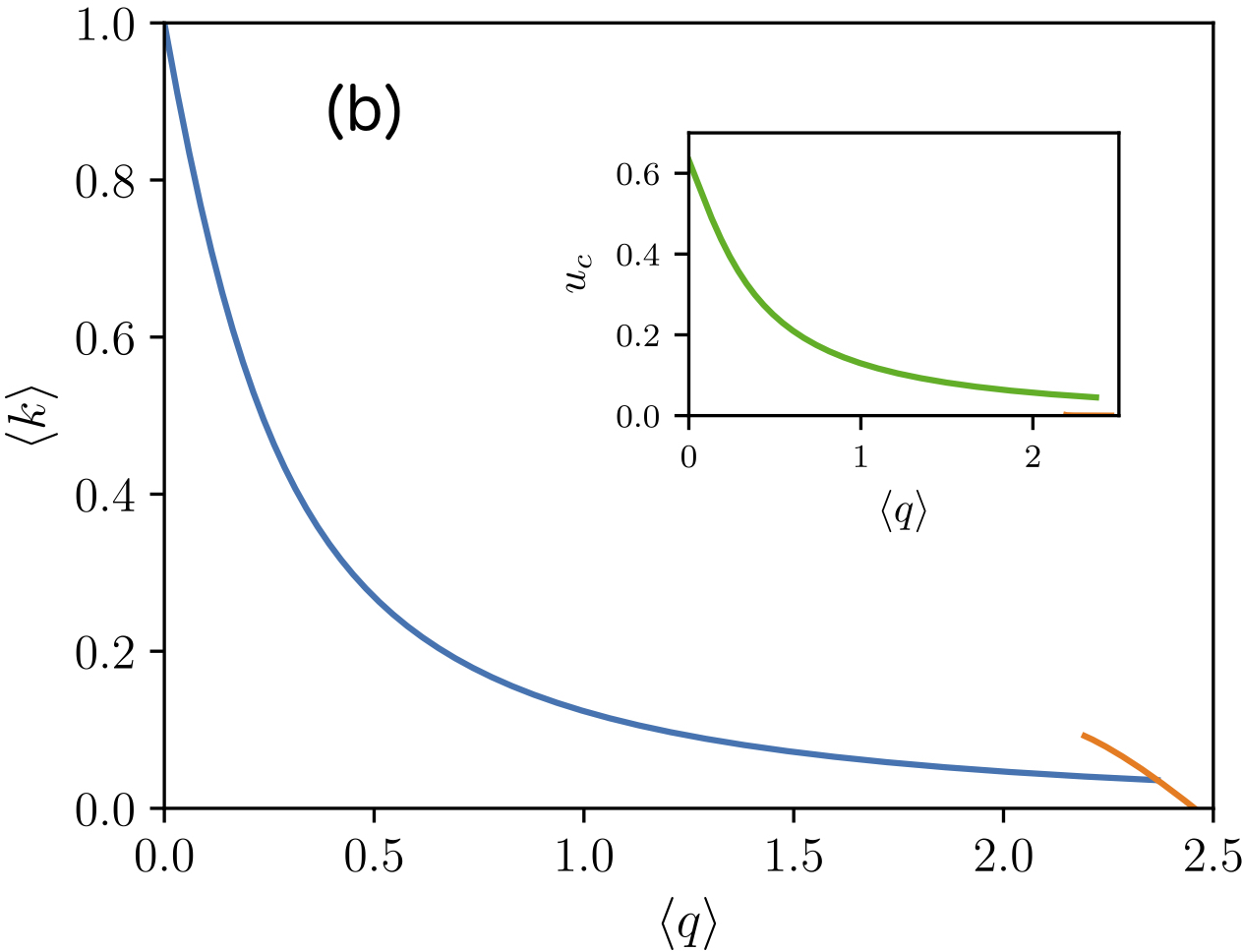} \\[2pt]
\includegraphics[width=0.8\columnwidth]{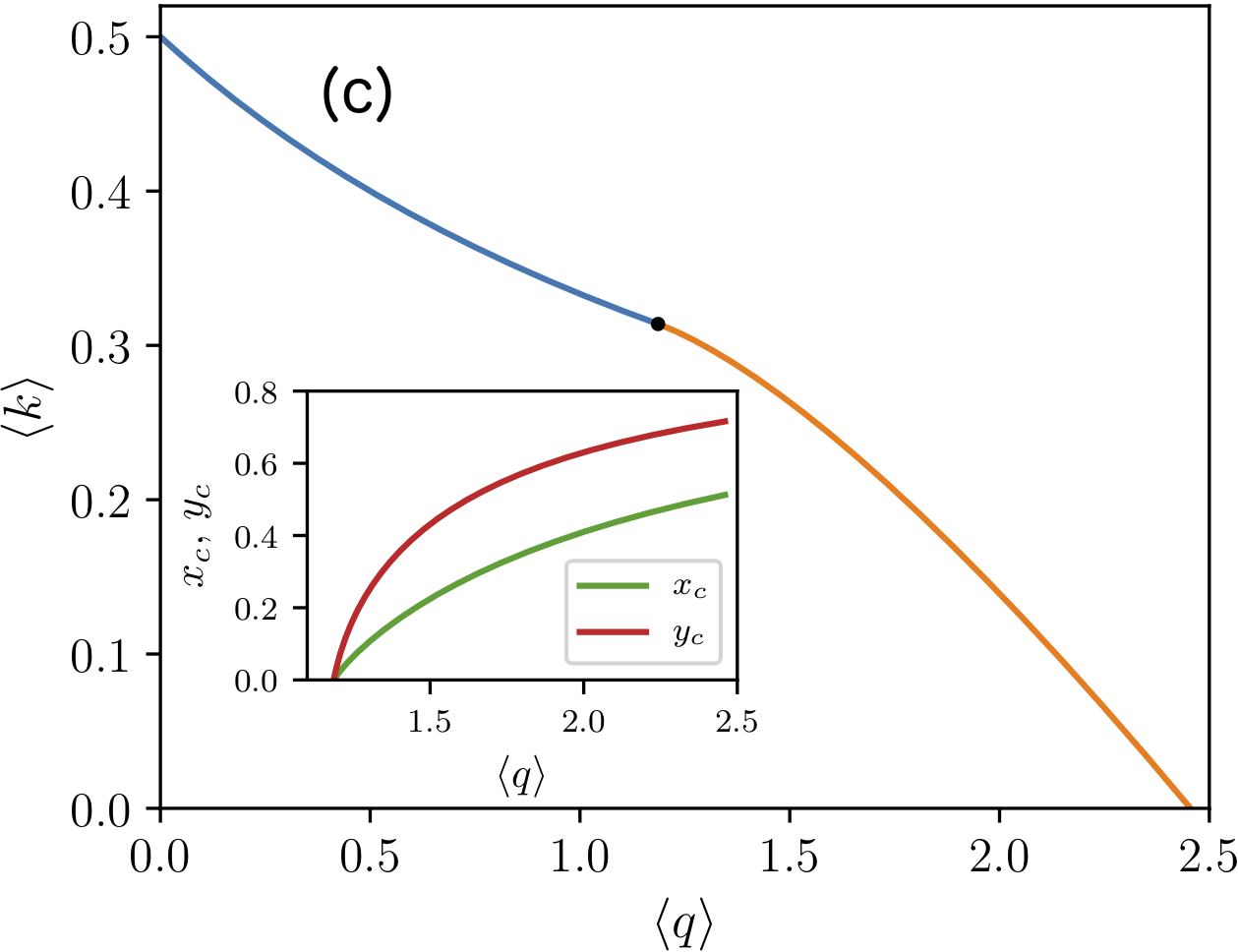} 
\end{center}
\caption{
(a) Phase diagrams in two layers with Poisson degree distributions (blue curve).  The inset shows the corresponding dependence of the finite branch probability $u_c$ on $\langle q \rangle$ (green curve). 
(b, c) Phase diagrams for three symmetric Poisson layers, with triple overlaps and double overlaps, respectively. 
Along the blue line the transition is continuous, and in the orange line it is discontinuous.
In the double overlap case, the two lines meet at a tricritical point $\langle q \rangle_\text{T}=(\sqrt{33}-1)/4$ and $\langle k \rangle_\text{T}=(7-\sqrt{33})/4$, marked with a black dot.
The inset in (b) represents the critical value of $u$ at the continuous transition.
The inset in (c) represents the size of the jump of probabilities $x$ and $y$ at the transition, which approach $0$ at the tricritical point. See Fig. \ref{f13} for the explanation of the probabilities $u$, $x$, and $y$.
}
\label{phase_diags}       
\end{figure}

This problem has not been considered at all in the weak percolation version of the problem.
In this paper we explore weak percolation in multilayer networks with overlaps.  
We develop a theory which enables us to obtain the phase diagram, which is significantly more rich than in the giant mutually connected component problem with overlaps, and to describe the phase transition associated with the birth of a giant component in this system. We show that the overlaps in our problem produce the opposite effect to that on the giant mutually connected component in interdependent networks. 
Namely, any (nonzero) concentration of overlaps drives the weak percolation transition to the ordinary percolation universality class in two layers, Fig.~\ref{phase_diags} (a),
and in three layers, we observe either a continuous or a discontinuous transition depending on the density of overlaps. This contrasts with interdependent networks where the overlaps fail to destroy the discontinuous transition. 

When overlaps occur only between two layers at once, we verify the existence of a tricritical point at which the lines of continuous and discontinuous transitions meet, Fig.~\ref{phase_diags} (c) . With triple overlaps, the lines do not meet at a tricritical point, Fig.~\ref{phase_diags} (b).

We consider first a two layer multiplex, in which there is only one kind of overlapped edge, allowing a compact formulation of the theory. The transition is always continuous.
We then extend these methods to consider three layer multiplexes, in which there are multiple combinations of edges that may overlap, and a discontinuous hybrid transition is present. We consider two representative particular cases, namely overlaps only between two layers at a time, and overlaps between all three layers (without any double overlaps).

The paper is structured as follows.
In Sec.~\ref{s2} we present and explain equations and the expression for the relative size of the giant component in two layers. 
In Sec.~\ref{s3} we analyze them and obtain the phase diagram and critical behavior.  
In Sec. \ref{sa2} we extend our analysis to consider three layers, first considering only triple overlaps. We then consider the case of double overlaps in Sec.~\ref{sa3}. 
In Sec.~\ref{s5} we discuss our results. 
In the Appendices we strictly derive the equations.


\section{Equations for two layers}
\label{s2}

\begin{figure}
\begin{center}
\includegraphics[width=0.9\columnwidth]{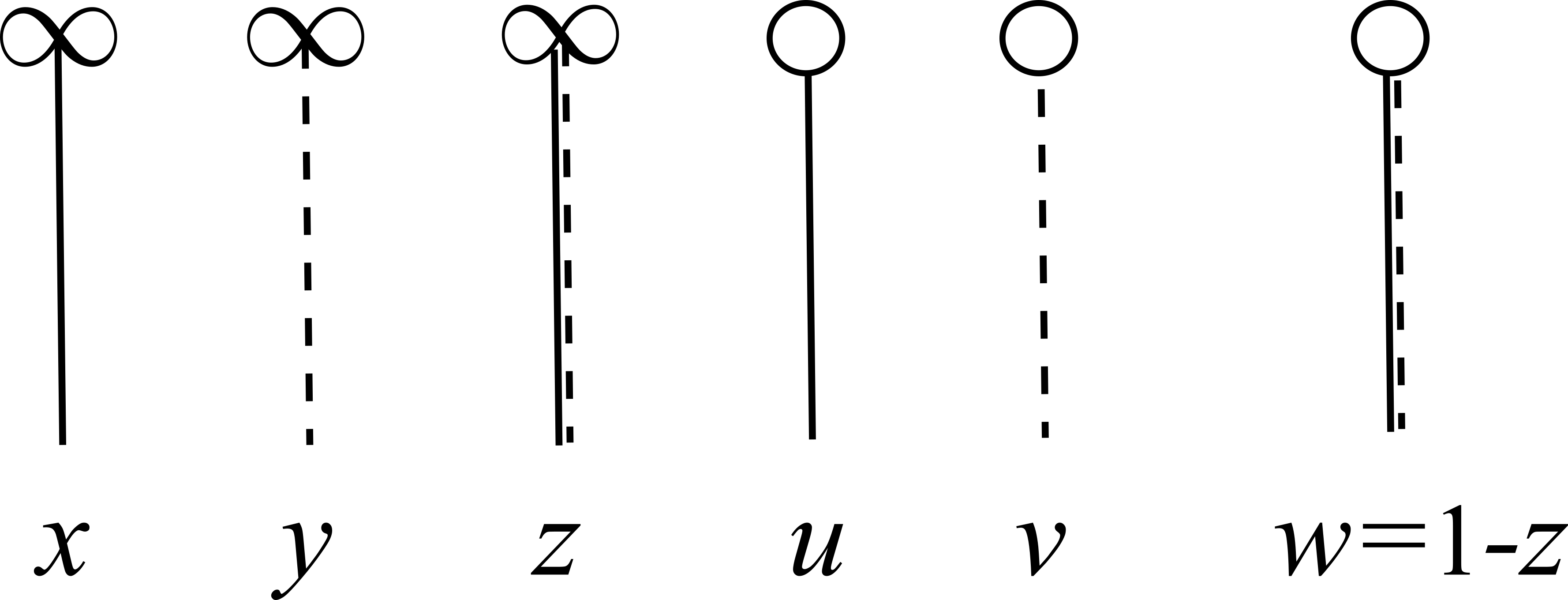}
\end{center}
\caption{
Probabilities (notations) for the weak multiplex percolation problem with overlaps, in two layers. 
The variables
$x$, $y$, and $z$ are the probabilities that edges of the corresponding type lead to infinite branches consisting of vertices satisfying the weak percolation condition. 
Conversely, 
$u$ and $v$ are the probabilities that edges of the corresponding type lead to finite weak branches, i.e., that all dead ends in these weak trees are edges with overlaps. For a double edge, the probability, $w$,  that it leads to a finite weak component equals $1-z$. 
}
\label{f13}       
\end{figure}

A multiplex network consists of a set of vertices with edges between them in $M$ layers. That is, each layer corresponds to a different type of edge. When two vertices are linked in more than one layer, we call this an overlapped connection, while if they are connected in only one layer, we call this a single edge.
For orientation and compactness of the analysis, we begin by describing the effect of overlaps in a two-layer multiplex.

Let us consider an infinite sparse random two-layer multiplex network, consisting of a set of nodes, each pair of which may be connected in one layer, the other layer or in both layers simultaneously. We refer to this last possibility as an overlapped or double edge. We can interpret these three possibilities as three different kinds of connections.

Generalizing the configuration model, we describe our infinite two-layer multiplex network by the joint degree--degree--degree distribution $P(q,q',k)$, where $q$ is the degree of a vertex in the first layer, $q'$ is its degree in the second layer and $k$ is its number of overlapped connections. The correlations of the three degrees of a vertex are the only correlations presenting in this network, and so the joint distribution $P(q,q',k)$ completely describes its structure. 

In the large size limit, such a multiplex is locally tree like, meaning that the probability to encounter certain configurations upon following an edge can be considered independent of the corresponding probabilities for other edges emanating from the same node. This allows us to write self-consistency equations for the probabilities of encountering specific configurations upon following a randomly chosen edge of a given type, whose solutions in turn allow us to calculate the size of the weak percolaton giant component.

In the weak percolation problem with overlaps, we introduce five probabilities, $x$, $y$, $z$, $u$, and $v$, see Fig.~\ref{f13}. 
The sixth probability, $w$, that a double edge leads to a finite weak component equals $1-z$.

The five probabilities can be found via the solution of five coupled equations.
We present the strict derivation of these equations in 
 \ref{sa1}, but one can understand them without derivation, see below. 
 
\begin{widetext}

Let us immediately list these equations: 
\begin{eqnarray}
&&
1 {-} x = \sum_{q,q',k} \frac{q}{\langle q \rangle}P(q,q',k)\bigl\{(1{-}x)^{q-1}(1{-}y)^{q'} (1{-}z)^k
+ \delta_{k,0}[1 - (1{-}x)^{q-1}](1{-}y{-}v)^{q'} \bigr\}
,
\label{1196}
\\[3pt]
&&
1 {-} y = \sum_{q,q',k} \frac{q'}{\langle q' \rangle}P(q,q',k)\bigl\{
(1{-}x)^q (1{-}y)^{q'-1} (1{-}z)^k + \delta_{k,0} (1{-}x{-}u)^q [1 - (1{-}y)^{q'-1}]
\bigr\}
,
\label{1197}
\\[3pt]
&&
1 {-} z = \sum_{q,q',k}\!\! \frac{k}{\langle k \rangle}P(q,q',k) (1{-}x)^q(1{-}y)^{q'}(1{-}z)^{k-1}
,
\label{1198}
\\[3pt]
&&
1 {-} x {-} u = \sum_{q,q'} \frac{q}{\langle q \rangle}P(q,q',0) (1{-}y{-}v)^{q'} 
,
\label{1199}
\\[3pt]
&&
1 {-} y {-} v = \sum_{q,q'} \frac{q'}{\langle q' \rangle}P(q,q',0) (1{-}x{-}u)^q 
.
\label{1200}
\end{eqnarray}
The solution of these equations should then be substituted into the following expression for the fraction $1-S$ of vertices outside of the giant component ($S$ is its relative size): 
\begin{eqnarray}
&&
\!\!\!\!\!\!\!\!\!\!\!\!
1 - S = 
\sum_{q,q',k} P(q,q',k)\Bigl\{
(1{-}x)^q (1{-}y)^{q'} (1{-}z)^k 
+ \delta_{k,0}\bigl\{
[1 {-} (1{-}x)^q](1{-}y{-}v)^{q'} +  (1{-}x{-}u)^q[1 {-} (1{-}y)^{q'}] 
\bigr\}
\Bigr\}
.
\label{1210}
\end{eqnarray}
We also strictly derive this expression in 
 \ref{sa1}.

\end{widetext}

One can understand the form of Eqs.~(\ref{1196})--(\ref{1210}) in the following way: 

Eq.~(\ref{1196}) is the probability that an edge of the first type leads only to finite weak branches.
The right-hand side of this equation has two terms, corresponding to the two ways that this can happen. The first is simply that all of the ongoing connections of all types lead to finite branches. The second possibility is that a connection to the giant component indeed exists, but that the node itself fails the weak percolation condition. For this to be the case, no overlapped edges can be present, as a single such edge satisfies the condition. Futhermore, the node must not have any connections to weak percolating branches via the other layer, whether finite or infinite (probabilities $u$ and $y$ respectively). 
Eq.~(\ref{1197}) can be understood in the same way,  simply exchanging probabilities for the first layer for those for the second layer.

Note that we must treat the case $k=0$ separately from the case $k \geq 1$, as evidenced by the presence of $\delta_{k,0}$ in these equations. This is because a single overlapped edge is sufficient to satisfy the weak percolation rule. This is not the case for the interdependent network problem, in which the connection to the giant component must be made in both layers, so terms like the second terms in Eqs. (\ref{1196}) and ~(\ref{1197}) are absent. This difference plays an important role in our analysis.

The right-hand side of Eqs.~(\ref{1198}) is straightforwardly the probability that all of the ongoing connections lead to only finite components.

Eq.~(\ref{1199}) is for the probability that an edge in the first layer leads to a node that fails the weak percolation condition. This can only be the case if $k=0$, and if any connections in the other layer do not lead to weakly percolating nodes. Eq.~(\ref{1200}) is for the reciprocal probability in the second layer.

Finally, the right-hand side of Eq.~(\ref{1210}) contains a term for the probability that none of the node's connections leads to the giant component, as well as terms for the futher possibilities that, in the case that $k=0$,  the node has connections to the giant component in one layer, or the other, but not both. 

In the absence of overlaps, Eqs.~(\ref{1196})--(\ref{1210}) reduce to the known equations for ``pure'' weak percolation \cite{baxter2021weak}. In the absence of single edges, Eqs.~(\ref{1196})--(\ref{1210}) are reduced to the standard equations for percolation \cite{dorogovtsev2008critical}.

Thus to solve our problem we should do the following: 

(i) solve Eqs.~(\ref{1199}) and (\ref{1200});  

(ii) substitute their solutions into Eq.~(\ref{1196})---(\ref{1198}) and solve them; 

(iii) substitute the complete set of these solutions into the expression for the size of the giant component, Eq.~(\ref{1210}).


\section{Phase diagram for two layers}
\label{s3} 

Let us use these equations to find the critical point at which a giant weak percolating cluster emerges.

At the critical point,  
$x=y=z=0$, assuming that the transition is continuous (which is the case, as we shall see later).   
Then Eqs.~(\ref{1199}) and (\ref{1200}) lead to the following two equations for the critical values of $u$ and $v$:
\begin{align}
1-u_c &= \sum_{q,q',k} \frac{q}{\langle q \rangle}P(q,q',0) (1{-}v_c)^{q'} 
,
\nonumber
\\
1-v_c &=  \sum_{q,q',k} \frac{q'}{\langle q' \rangle}P(q,q',0) (1{-}u_c)^q 
,
\label{1230}
\end{align}
In general, $0<u_c,v_c\leq1$, and in particular, if $P(q,q',k{=}0)=0$, then $u_c=v_c=1$, while $u=v=0$ if overlaps are absent, see below. 

Assuming
that the involved moments are finite, and linearizing Eq.~(\ref{1200}) for small $x$, $y$ ,$z$, and for small $\mu\equiv u-u_c$, and $\nu\equiv v-v_c$, we obtain
\begin{align}
x &= \begin{multlined}[t] \frac{1}{\langle q \rangle} \Bigl\{  
\bigl[ \langle q(q{-}1) \rangle - \langle q(q{-}1)(1{-}v_c)^{q'} \rangle_0 \bigr] x
\\ + \langle qq' \rangle y + \langle qk \rangle z 
\Bigr\}, \end{multlined}
\nonumber
\\
y &= \begin{multlined}[t] \frac{1}{\langle q' \rangle}
\Bigl\{ 
\langle qq' \rangle x \\ + \bigl[ \langle q'(q'{-}1) \rangle 
- \langle q'(q'{-}1)(1{-}u_c)^{q} \rangle_0 \bigr] y + \langle q'k \rangle z  
\Bigr\}, \end{multlined}
\nonumber
\\[3pt]
z &= \frac{1}{\langle k \rangle} 
\Bigl\{ \langle qk \rangle x + \langle q'k \rangle y + \langle k(k{-}1) \rangle z \Bigr\}
. 
\label{1240}
\end{align}
Here 
\begin{align}
\langle q(q{-}1)(1{-}v_c)^{q'} \rangle_0 &\equiv \sum_{q,q'} q(q{-}1)(1{-}v_c)^{q'}\! P(q,q',0)
,  
\nonumber
\\\langle q'(q'{-}1)(1{-}u_c)^q \rangle_0 &\equiv \sum_{q,q'} q'(q'{-}1)(1{-}u_c)^q P(q,q',0)
.
\label{1245}
\end{align}
Note that the linearized equations for $x,y,x$ actually do not contain $\mu$ and $\nu$ (they contain only $v_c$ and $u_c$), and so they are separated from the equations for $u,v$. (Expanding the right-hand sides of the first two equations of  Eq.~(\ref{1200}) already gives nonlinear terms $x\nu$ and $y\mu$, which should be ignored.) 
Note that in the particular case of the pure weak percolation problem, $u_c=v_c=0$, and so the coefficients $x$ and $y$ in the first and second equations, respectively disappear as they should. 

\begin{widetext}

The condition for the phase transition surface is then
\begin{equation}\text{det}\!
\left( 
\begin{array}{ccc}
 \langle q(q-2) \rangle {-} \langle q(q{-}1)(1{-}v_c)^{q'} \rangle_0 \ \ & \langle qq' \rangle & \langle qk \rangle
\\[3pt]
\langle qq' \rangle \ \ & \langle q'(q'-2) \rangle {-} \langle q'(q'{-}1)(1{-}u_c)^q \rangle_0 \ \ & \langle q'k \rangle  
\\[3pt]
\langle qk \rangle \ \ & \langle q'k \rangle & \langle k(k-2) \rangle
\end{array}
\right) = 0 
,
\label{1250}
\end{equation}
where $u_c$ and $v_c$ are the solutions of Eq.~(\ref{1230}). 
In the particular case of $P(q,q',k>0)=0$ we immediately get the following threshold for pure weak percolation (that is, weak percolation without overlapped edges): 
\begin{equation}
\langle q \rangle \langle q' \rangle = \langle qq' \rangle^2
.  
\label{1280}
\end{equation}

\end{widetext}

\subsection{Symmetric uncorrelated situation}

Let us consider the simplifying case of uncorrelated degrees, so that $P(q,q',k) = P(q)P(q')Q(k)$. 
Then $x=y$ and $u=v$. So the equations are 
\begin{align}
x &= \begin{multlined}[t]
1 {-} \frac{G'(1{-}x)}{\langle q \rangle}G(1{-}x)R(1{-}z) {-} 
\\
[1 {-} \frac{G'(1{-}x)}{\langle q \rangle}]
(1{-}x{-}u)  
,\end{multlined}
\nonumber
\\
z &= 1 - G^2(1{-}x) \frac{R'(1{-}z)}{\langle k \rangle} 
,
\nonumber
\\
1{-}x{-}u &= Q(0) G(1{-}x{-}u)
,   
\label{1370}
\end{align}
and the expression for $S$ is
\begin{equation}
S = 1 - G^2(1{-}x)R(1{-}z) - 2
[1 {-} G(1{-}x)]
(1{-}x{-}u),
\label{1380}
\end{equation}
where we have written the expressions in terms of the generating functions $G(z) = \sum_q P(q)z^q$ and  $R(z) = \sum_k Q(k)z^k$ of the single and overlapped edge degree distributions, respectively.

Then Eq.~(\ref{1250}) is reduced to 
\begin{equation}
\text{det}\!
\left( \!\!
\begin{array}{ccc}
 u_c\langle q(q{-}1) \rangle {-} \langle q \rangle\ \ & \langle q \rangle^2 & \langle q \rangle \langle k \rangle
\\[3pt]
\langle q \rangle^2 \ \ & u_c\langle q(q{-}1) \rangle {-} \langle q \rangle \ \ & \langle q \rangle \langle k \rangle 
\\[3pt]
\langle q \rangle \langle k \rangle \ \ & \langle q \rangle \langle k \rangle & \langle k(k{-}2) \rangle
\end{array}\!\!
\right) = 0 
,  
\label{1290}
\end{equation}
where 
\begin{equation}
1 {-} u_c = Q(0) G(1{-}u_c) 
 = Q(0)\langle (1{-}u_c)^q \rangle 
,  
\label{1300}
\end{equation}
i.e., $u_c= F[ P(q), Q(0) ]$, a functional. 
The solution $u_c$ of this equation behaves in the following way. $u_c$ increases from $1-Q(0)$ to $1$ as $\langle q \rangle$ increases from $0$ to $\infty$ [one can check this by considering a Poisson distribution $P(k)$]. For Poisson distributions, if $Q(0)=1$, then $u_c=0$ for $0<\langle q \rangle<1$,  and $u_c$ increases from $0$ to $1$ as $\langle q \rangle$ increases from $1$ to $\infty$. Apparently, the critical point in this situation should be certainly below the critical point without overlaps, which is $\langle q \rangle=1$, so we have $u_c=0$ if $Q(0)=1$.  

Thus the equation for the critical line is 
\begin{multline}
\bigl[ u_c\langle q(q{-}1) \rangle - \langle q \rangle- \langle q \rangle^2 \bigr] 
\\
\times
 \bigl\{ \bigl[ u_c\langle q(q{-}1) \rangle - \langle q \rangle + \langle q \rangle^2 \bigr] \langle k(k{-}2) \rangle - 2 \langle q \rangle^2 \langle k \rangle^2 \bigr\} 
\\
= 0
.  
\label{1310}
\end{multline}

This equation provides two possible branches: 
\begin{align}
u_c\langle q(q{-}1) \rangle - \langle q \rangle - \langle q \rangle^2  
&= 0
,
\nonumber
\\
\bigl[ u_c\langle q(q{-}1) \rangle - \langle q \rangle + \langle q \rangle^2 \bigr] \langle k(k-2) \rangle 
- 2 \langle q \rangle^2 \langle k \rangle^2  
&= 0
.  
\label{1320}
\end{align}
If both these branches are indeed present, they would meet 
at the point where the following necessary condition is satisfied:  
\begin{equation} 
\langle k(k-2) \rangle =  \langle k \rangle^2 
.  
\label{1330}
\end{equation}
This corresponds to a more dense network than the ordinary percolation threshold (at which $\langle k(k-2) \rangle = 0$) of the network with the same degree distribution of overlaps but without single edges. The additional single edges should in fact decrease the threshold.  Consequently this crossing is impossible.

Note that in the case of $\langle q \rangle \to 0$, the second equation in Eq.~(\ref{1320}) gives $\langle k(k-2) \rangle=0$ as it should be. Indeed, in this situation we have $1-u_c \to Q(0)$, $\langle q^2 \rangle \cong \langle q \rangle \ll1$, so $\bigl[ -\langle q \rangle {-} Q(0) (\langle q \rangle{-}\langle q \rangle) + \langle q \rangle^2 \bigr] \langle k(k-2) \rangle - 2 \langle q \rangle^2 \langle k \rangle^2  
= 0$. This directly leads to $\langle k(k-2) \rangle=0$.

Furthermore, in the particular case of $\langle k \rangle=0$, the second equation in Eq.~(\ref{1320}) gives $\langle q \rangle=1$ as it should be for the critical point without overlaps. Indeed, consider $\langle k \rangle\to0$ and $Q(0)\to1$. We take into account that $u_c=0$ in this case, see above. 
Let us assume that only $Q(0)$ and $Q(1)$ are non-zero, and $Q(1)\ll1$, then $\langle k \rangle^2 \cong \langle k \rangle\ll1$, and so we have $[-\langle q \rangle + \langle q \rangle^2](-\langle k \rangle) - 2 \langle q \rangle^2 \langle k \rangle^2  = 0$. This leads to $\langle q \rangle=1$.

Thus both limiting cases, namely, pure weak percolation (no overlaps) and ordinary percolation (only overlaps), are described by the second equation in Eq.~(\ref{1320}) and not the first one.

We get the final equations for these two branches by substituting the solution of Eq.~(\ref{1300}), $1-u_c=1-F[ P(q), Q(0) ]$,  into Eq.~(\ref{1320}), providing the condition containing $P(q)$, $Q(0)$, $\langle k \rangle$, and $ \langle k^2 \rangle$. This is certainly possible only if the right-hand sides of the equations 
\begin{align}
1{-}u_c &= \frac{\langle q(q-2) \rangle  - \langle q \rangle^2}{ \langle q(q{-}1) \rangle}  ,
\nonumber
\\[3pt]
1{-}u_c &= 
\frac{\bigl[\langle q(q-2) \rangle  + \langle q \rangle^2\bigr] \langle k(k-2) \rangle  - 2 \langle q \rangle^2 \langle k \rangle^2 }{\langle q(q{-}1) \rangle \langle k(k-2) \rangle}
\label{1335}
\end{align}
are positive, as $u_c$ is a probability. For the first equation in Eq.~(\ref{1335}), this is possible if 
\begin{equation} 
\langle q(q-1) \rangle > \langle q \rangle^2 + \langle q \rangle 
,  
\label{1337}
\end{equation}
which is not the case for \ER networks. 
On the other hand, the exponentially decreasing $P(q)=2^{-1-q}$ gives $\langle q(q-1) \rangle = \langle q \rangle^2 + \langle q \rangle$.

We now give the solutions for these two representative cases.

\subsubsection{The particular case of  \ER graphs}

In the case of Poisson distributions $P(q)$ and $Q(k)$, that is, \ER layers,
$\langle q(q-1) \rangle = \langle q \rangle^2$, 
$\langle k(k-1) \rangle = \langle k \rangle^2$, and $Q(0)=e^{-\langle k \rangle}$,
while the generating functions may be written $G(x) = e^{-\langle q \rangle(1-x)}$, 
and $R(x) = e^{-\langle k \rangle(1-x)}$, so we have, from Eqs. (\ref{1370}) and (\ref{1380}),
\begin{align}
x &= z - [1 - e^{-\langle q \rangle x}](1{-}x{-}u)  
,
\nonumber
\\[3pt]
z &= 1 - e^{-2\langle q \rangle x} e^{-\langle k \rangle z} 
,
\nonumber
\\[3pt]
1{-}x{-}u &= e^{-\langle k \rangle} e^{-\langle q \rangle(x+u)}
,   
\label{1410}
\end{align}
and 
\begin{equation}
S = 2x-z.  
\label{1420}
\end{equation}

Hence we have the two branches
\begin{align}
\langle q \rangle^2 - \langle q \rangle - (1{-}u_c) \langle q \rangle^2 - \langle q \rangle^2  
&= 0
,
\nonumber
\\[3pt]
\bigl[ \langle q \rangle^2 {-} \langle q \rangle {-} (1{-}u_c) \langle q \rangle^2 {+} \langle q \rangle^2 \bigr] (\langle k \rangle^2 {-} \langle k \rangle) {-} 2 \langle q \rangle^2 \langle k \rangle^2  
&= 0
,  
\label{1340}
\end{align}
and, from Eq.~(\ref{1330}), we obtain
\begin{equation} 
\langle k \rangle^2 - \langle k \rangle  =  \langle k \rangle^2 
,  
\label{1350}
\end{equation}
that is, there is no crossing. 
Indeed, for this \ER case, the first equations in Eqs.~(\ref{1320}) or (\ref{1335}) certainly have no real solutions, so this branch is absent in this case.

Keeping only the second branch, we can obtain the critical line in terms of $\langle q \rangle$ and $\langle k \rangle$ from the following two equations: 
\begin{align}
\bigl[ 2\langle q \rangle - (1-u_c) \langle q \rangle - 1  \bigr] (\langle k \rangle - 1) 
&= 2 \langle q \rangle \langle k \rangle
,
\nonumber
\\[3pt]
1-u_c &= e^{-\langle k \rangle} e^{-\langle q \rangle u_c}
.  
\label{1360}
\end{align}
The critical line obtained from numerical solution of these equations is shown in Fig.~\ref{phase_diags}(a).

\begin{figure}
\begin{center}
\includegraphics[width=\columnwidth]{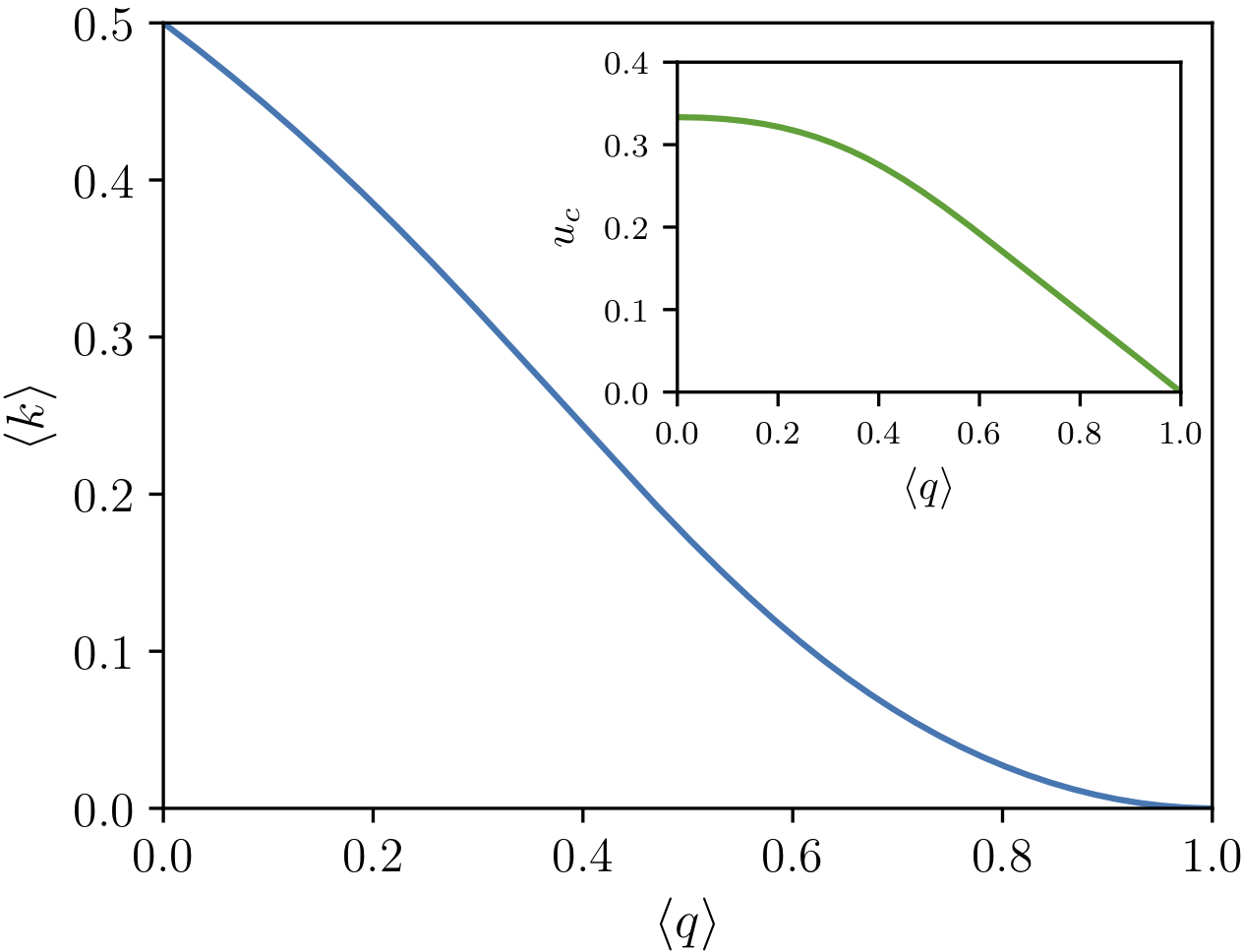}
\end{center}
\caption{
Phase diagrams in two layers with exponential degree distributions (blue curves online). Only the lower branch is realized, and the real branch of the phase border and the non-physical line do not touch each other. The inset shows the corresponding dependence of $u_c$ on $\langle q \rangle$ (green curve).
}
\label{f14}       
\end{figure}

When $\langle q \rangle = 0$ and $\langle k \rangle = 0$ these equations and expression are reduced to ordinary percolation and pure weak percolation, respectively. 

\medskip

\subsubsection{Exponential degree distributions}

Let us now consider another example, exponential degree distributions $P(q)= \dfrac{c{-}1}{c}\,c^{-q}$, $Q(k) = \dfrac{d{-}1}{d}\,d^{-k}$, where $c,d>1$. 
Then the first equation has a solution, which is, however, non-physical, since this branch is already in the phase with a giant component, where we cannot set $x=y=z=0$. The shapes of the second (physical) branch is shown in Fig.~\ref{f14}. Compare with the phase diagrams for \ER layers, Fig. \ref{phase_diags} (a).

Hence in both these examples, we find that the lower branch is the physical one, and observe qualitatively similar phase diagrams.


\section{Three-layer multiplex with only triple overlaps}
\label{sa2}

We now extend our analysis to multiplex networks with three layers. In the pure weak percolation problem a discontinuous hybrid phase transition appears when there are three or more layers \cite{baxter2021weak}.  
There are now multiple possible types of overlaps, as edges may overlap in two layers or in all three layers.
Here we consider two representative cases: only single and triply overlapped edges, and only single and doubly overlapped edges.

Let a $3$-layer multiplex have only triply overlapping edges---only a triple of edges taken different layers can form an overlap. So we consider a maximally random multiplex with a given joint distribution $P(q,q',q'',k)$, where $q$, $q'$, and $q''$ are degrees for non-overlapping edges within three layers, and $k$ are vertex degrees counting the triple overlaps.

In this system, we need to find seven probabilities corresponding to the probabilities for single edges to lead to infinite sub-trees ($x$, $y$ and $z$) or finite sub-trees ($u$, $v$, and $w$) as well as the probability $r$ that a triple overlapped edge leads to an infinite sub-tree.

\begin{widetext}
There are seven equations for the probabilities $x$, $y$, $z$, $r$, $u$, $v$, and $w$, see 
 \ref{sa33} for their derivation. Firstly, 
\begin{multline}
1 {-} x = \sum_{q,q',q'',k} \frac{q}{\langle q \rangle}P(q,q',q'',k) 
\Bigl\{ (1{-}x)^{q-1} (1{-}y)^{q'} (1{-}z)^{q''} (1{-}r)^k 
\Bigr. 
\\
+ \delta_{k,0} \bigl\{ 
(1{-}y{-}v)^{q'}[ 1 - (1{-}x)^{q-1} (1{-}z)^{q''} ] + (1{-}z{-}w)^{q''} [ 1 - (1{-}x)^{q-1} (1{-}y)^{q'} ] - [ 1 - (1{-}x)^{q-1}] (1{-}y{-}v)^{q'} (1{-}z{-}w)^{q''} 
\\
+[1 - (1{-}r)^0][-(1{-}y)^{q'} - (1{-}z)^{q''} + (1{-}y)^{q'} (1{-}z)^{q''} + (1{-}x)^{q-1} (1{-}y)^{q'} (1{-}z)^{q''}]
\bigr\} 
\Bigr\}
,
\label{b100}
\end{multline}
while two more equations, for $1-y$ and $1-z$, have an identical structure and may be found by cycling the variables.
Then
\begin{align}
1 {-} r = \sum_{q,q',q'',k} \frac{k}{\langle k \rangle} P(q,q',q'',k) (1{-}x)^q (1{-}y)^{q'} (1{-}z)^{q''} (1{-}r)^{k-1}
,
\label{b130}
\end{align}
and
\begin{align}
1 {-} x {-} u = \sum_{q,q',q''} \frac{q}{\langle q \rangle}P(q,q',q'',0) 
[ (1{-}y{-}v)^{q'} + (1{-}z{-}w)^{q''} - (1{-}y{-}v)^{q'} (1{-}z{-}w)^{q''} ]
,
\label{b140}
\end{align}
and there are two further similar equations, for $1-y-v$ and $1-z-w$, which again may be found by appropriately cycling the variables.

The equation for the giant component size $S$ is then,
\begin{multline}
1 - S = 
\sum_{q,q',q'',k} P(q,q',q'',k)\Bigl\{
(1{-}x)^q (1{-}y)^{q'} (1{-}z)^{q''} (1{-}r)^k 
\Bigr. 
\\
\Bigl. 
+ \delta_{k,0}\bigl\{ 
(1{-}x{-}u)^q + (1{-}y{-}v)^{q'} \! + (1{-}z{-}w)^{q''} \! - (1{-}x{-}u)^q (1{-}y{-}v)^{q'} \! - (1{-}y{-}v)^{q'} (1{-}z{-}w)^{q''} \! - (1{-}x{-}u)^q (1{-}z{-}w)^{q''} 
\\
+ (1{-}x)^q (1{-}y{-}v)^{q'} (1{-}z{-}w)^{q''} 
+ (1{-}x{-}u)^q (1{-}y)^{q'} (1{-}z{-}w)^{q''}
+ (1{-}x{-}u)^q (1{-}y{-}v)^{q'} (1{-}z)^{q''} 
\\
- (1{-}x{-}u)^q(1{-}y)^{q'} (1{-}z)^{q''} - (1{-}x)^q (1{-}y{-}v)^{q'} (1{-}z)^{q''} - (1{-}x)^q (1{-}y)^{q'} (1{-}z{-}w)^{q''} 
\bigr\}
\Bigr\}
.
\label{b20}
\end{multline}

Let us once again consider uncorrelated layers. Let $P(q,q',q'',k) = P(q) P(q') P(q'') Q(k)$, and the generating functions of $P(q)$ and $Q(k)$ be $G(z)$ and $R(z)$, respectively.  This gives $x=y=z$, $r$, $u=v=w$. Then 
\begin{align}
1 {-} x &= %
\frac{G'(1{-}x)}{\langle q \rangle} [G(1{-}x)]^2 R(1{-}r)
+ Q(0) G(1{-}x{-}u) \!\biggl\{ 2 \biggl[1 - \frac{G'(1{-}x)}{\langle q \rangle} G(1{-}x) \biggr]
 - 
\!\biggl[ 1 - \frac{G'(1{-}x)}{\langle q \rangle} \biggr] G(1{-}x{-}u) 
\!\biggr\}
,
\label{b210}
\\
1 {-} r &= [G(1{-}x)]^3 \frac{R'(1{-}r)}{\langle k \rangle} 
,
\label{b220}
\\
1 {-} x {-} u &= Q(0) \left\{ 2 G(1{-}x{-}u)  - [G(1{-}x{-}u)]^2 
\right\} 
\label{b230}
\end{align}
and 
\begin{align}
1 - S = [G(1-x)]^3 R(1-r) 
+ 3 Q(0) G(1-x-u) 
\Big\{ 1 - G(1-x-u) 
+ G(1-x) G(1-x-u) - [G(1-x)]^2  \Big\}
.
\label{b240}
\end{align}
\end{widetext}

\begin{figure}
\begin{center}
\includegraphics[width=\columnwidth]{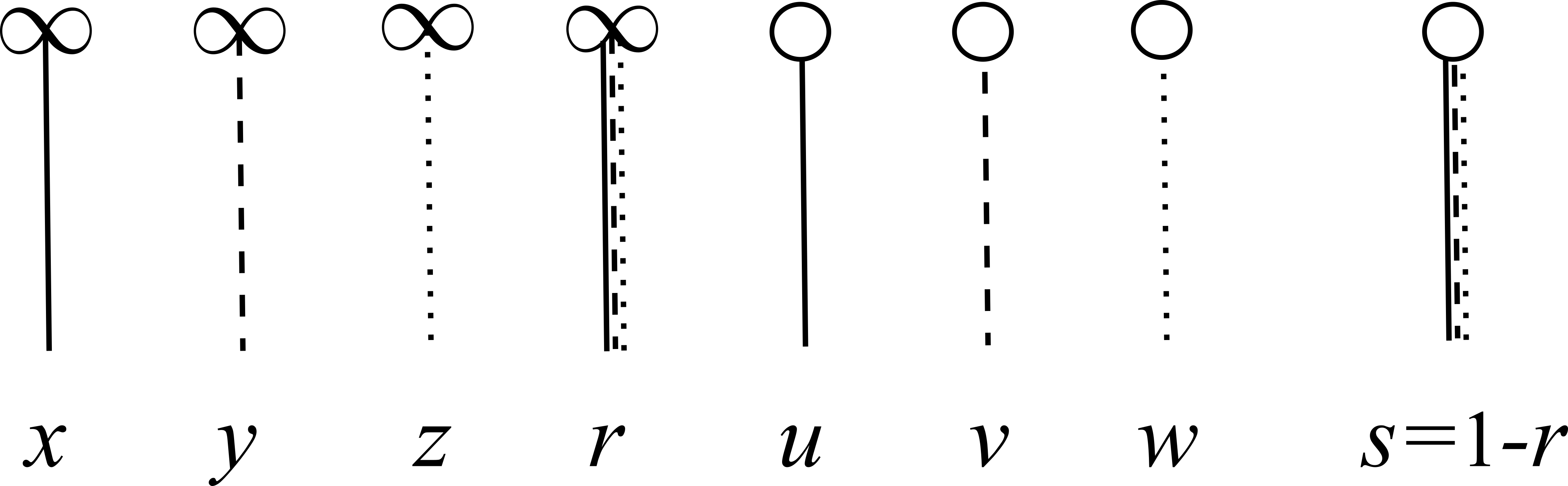}
\end{center}
\caption{
Probabilities (notations) for weak multiplex percolation in three layers with triple overlaps. 
The variables
$x$, $y$, $z$, and $r$ are the probabilities that edges of the corresponding type lead to infinite branches consisting of vertices satisfying the weak percolation condition. Conversely,
$u$ $v$, and $w$ are the probabilities that edges of the corresponding type lead to finite weak branches, i.e., that all dead ends in these weak trees are edges with overlaps. For a triple edge, the probability, $s$,  that it leads to a finite weak component equals $1-r$. 
}
\label{f5}       
\end{figure}

\subsection{Phase diagram for \ER layers}

To illustrate the critical behavior in this case, we now consider the specific example of \ER layers. The critical phenomena observed will be qualitatively the same for any degree distributions whose first and second moments remain finite in the infinite system size limit.
For Poisson distributions with the first moments $\langle q \rangle$ and $\langle k \rangle$, we have
\begin{align}
1 {-} x &= \begin{multlined}[t]
e^{-3\langle q \rangle x} e^{-\langle k \rangle r} 
\\
+ e^{-\langle k \rangle} e^{-\langle q \rangle (x+u)} 
\Big\{2\left[ 1 -  e^{-2\langle q \rangle x} \right] 
\\
- \left[ 1 -  e^{-\langle q \rangle x} \right] e^{-\langle q \rangle (x+u)} \Big\}
,\end{multlined}
\label{b250}
\\
1 - r &= e^{-3\langle q \rangle x} e^{-\langle k \rangle r} 
,
\label{b260}
\\
1 - x - u &= e^{-\langle k \rangle} \left\{ 2 e^{-\langle q \rangle (x+u)} - e^{-2\langle q \rangle (x+u)}  
\right\} 
\label{b270}
\end{align}
and 
\begin{multline}
1 - S = e^{-3\langle q \rangle x} e^{-\langle k \rangle r} 
\\
+ 3 e^{-\langle k \rangle} e^{-\langle q \rangle (x+u)} 
\Big[ 1 - e^{-\langle q \rangle (x+u)} 
\\
+ e^{-\langle q \rangle x} e^{-\langle q \rangle (x+u)} - e^{-2\langle q \rangle x}  \Big]
.
\label{b280}
\end{multline}

\begin{figure}[h]
\begin{center}
\includegraphics[width=\columnwidth]{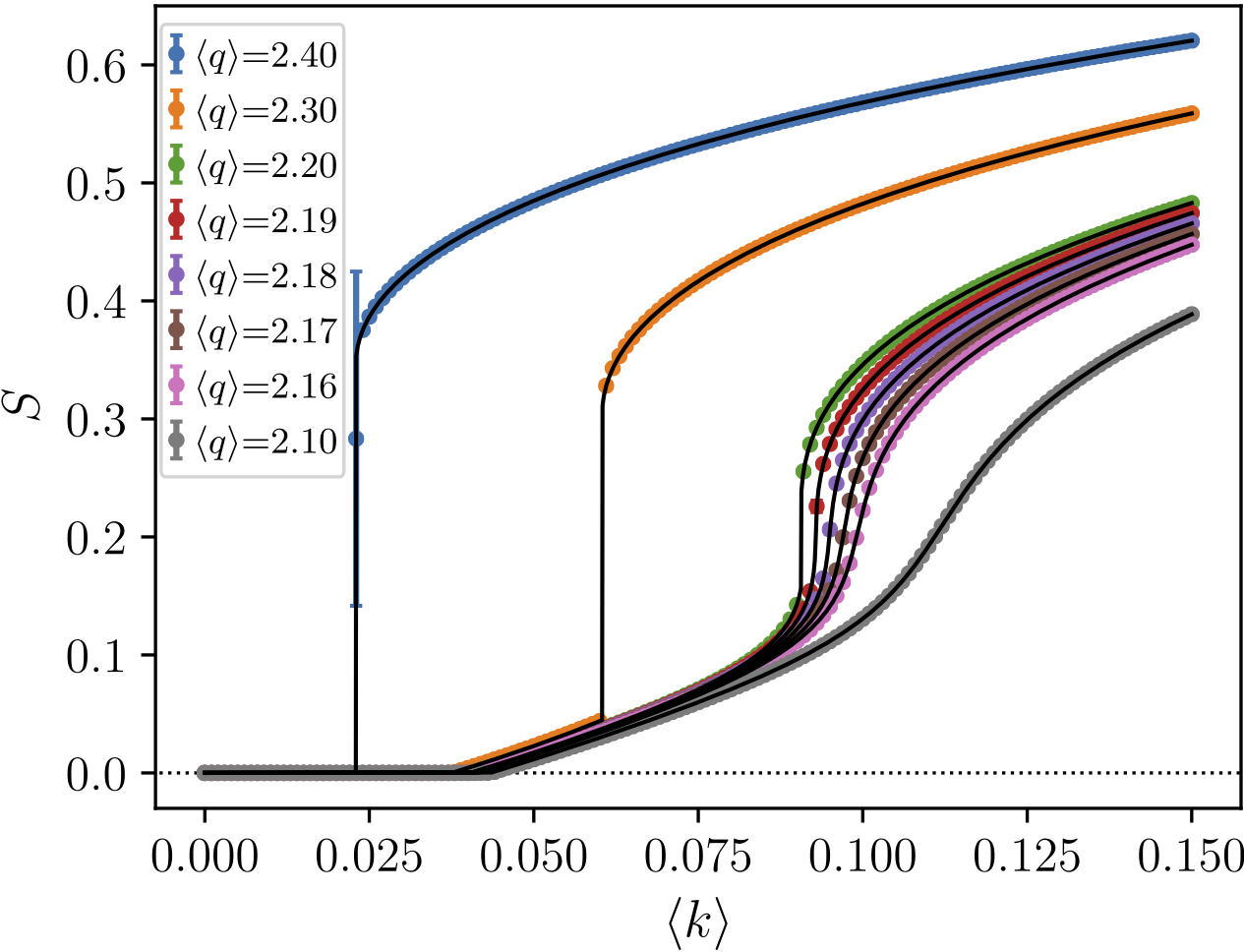} 
\end{center}
\caption{
Results of simulations of weak percolation in the presence of single and triply overlapped edges (no double edges), for symmetrical uncorrelated Poisson degree distributions with means $\langle q \rangle$ and $\langle k \rangle$, respectively. 
Network are generated with $10^8$ nodes, and each point is the average of $10$ realizations.
Black lines represent Eq.~(\ref{b280}) with $x$, $r$, and $u$ replaced by the solution of Eqs.~(\ref{b250})--(\ref{b270}).
}
\label{f17}       
\end{figure}

Numerical solutions of these equations for certain values of the single mean degree $\langle q\rangle$ between $2.1$ and $2.4$ are shown in Fig. \ref{f17}. We see that as the overlap mean degree $\langle k\rangle$ increases, we may encounter either a continuous transition, a continuous transition followed by a discontinuous one, or only a discontinuous transition, depending on the value of $\langle q\rangle$.

The phase diagram with the line of continuous transitions and line of discontinuous transitions is shown in Fig.~\ref{phase_diags} (b). 
Along the blue curve beginning from $\langle q\rangle = 0, \langle k\rangle = 1$, there is a continuous phase transition, with exponent $1$. That is, the transition is of the same universality class as ordinary percolation.  As we approach the continuous transition line, both $x$ and $r$ tend to 0, and $S \cong Ax+Br$. This can be seen by linearizing  Eq. (\ref{b280}). In the limit $\langle q \rangle \to 0$ the weak percolation rule reduces to percolation of the triple overlap edges. As we will see in the following section, the exponent becomes $2$ in the case of double overlaps.
For larger values of $\langle q\rangle$, and low concentrations of overlaps, the line of continuous transitions meets a second line of transitions (orange line in the figure), at which there is a jump in the size of the giant component. On the left-hand side to the point of meeting, this jump is from a nonzero value.
 To the right from the meeting point, the jump is from zero, see Fig. \ref{f17}. This discontinuous hybrid transition is also observed in the absence of overlaps.



\section{Three-layer multiplex with only double overlaps}
\label{sa3} 

Now let us consider the case of only double overlaps (i.e. no triple overlaps).
Let a $3$-layer multiplex have only single doubly overlapping edges---only a pair of edges taken two different layers can form an overlap. We consider a maximally random multiplex with a given joint distribution $P(q,q',q'',k, k', k'')$, where $q$, $q'$, and $q''$ are degrees for non-overlapping edges within three layers, and $k$, $k'$, and $k''$ are vertex degrees counting the double overlaps within each of three pairs of distinct layers.  

\begin{figure}[h]
\begin{center}
\includegraphics[width=0.8\columnwidth]{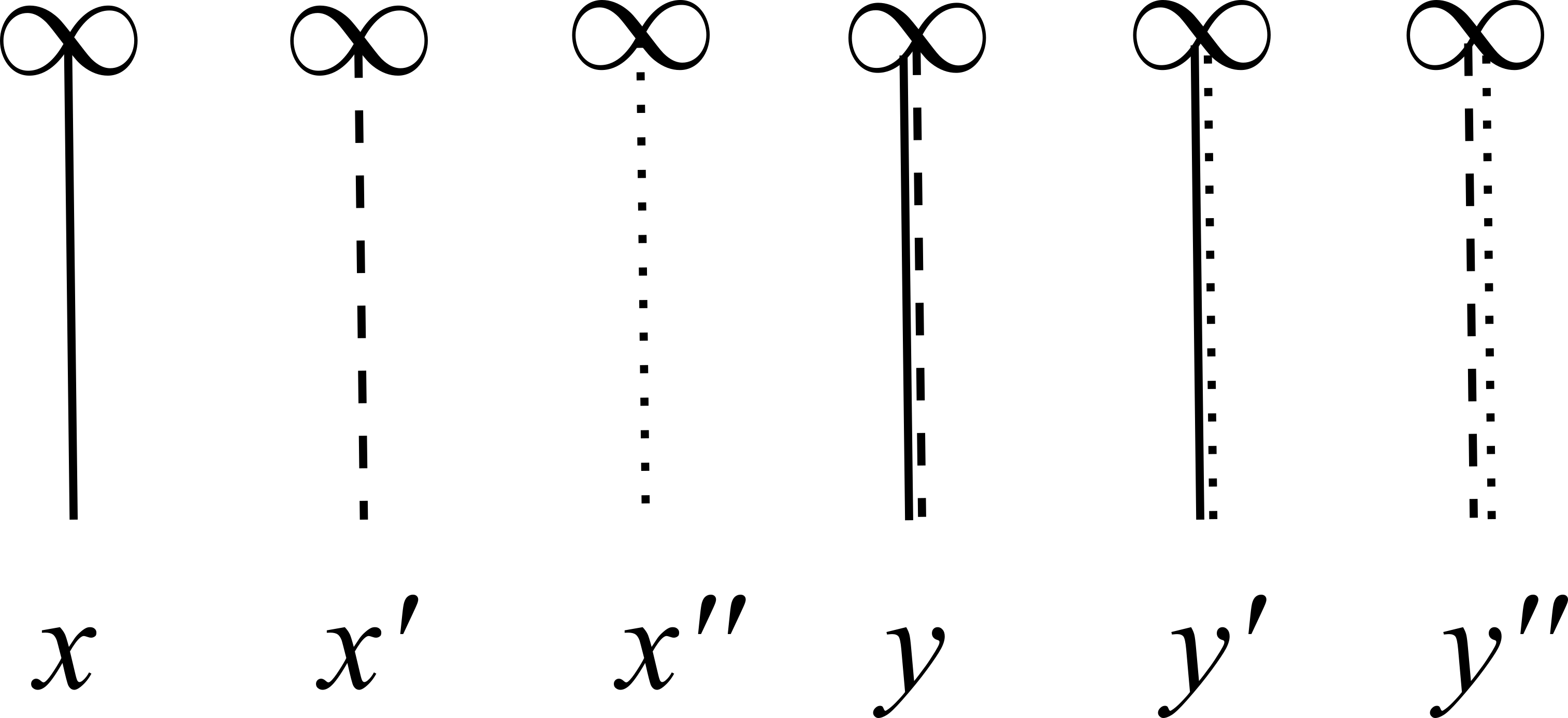}
\end{center}
\caption{
Probabilities (notations) for weak multiplex percolation in three layers with double overlaps. 
The variables
$x$, $x'$, and $x''$, are the probabilities that single edges of the corresponding type lead to infinite weak branches. Similarly, $y$, $y'$, and $y''$, are the probabilities that overlapped edges of the corresponding type lead to infinite weak branches. 
}
\label{f15}       
\end{figure}

In the absence of triple edges overlapped in all three layers there cannot be finite viable components, because the network is locally tree-like.
So, in this problem we need to consider six probabilities, $x$, $x'$, $x''$, $y$, $y'$, and $y''$, defined in Fig.~\ref{f15}, and six equations, as well as an expression for the relative size of the giant component $S$.
We derive these six equations and the expression for $S$ in 
 \ref{sa32}, following the approach introduced in Section~\ref{sa2} and 
  \ref{sa1}.
  
\begin{widetext}
The six self-consistency equations for the probabilities $x$,  $x'$,  $x''$,  $y$,  $y'$, and  $y''$ are
\begin{align}
1 {-} x  &= \begin{multlined}[t] 
\sum_{q,q',q'',k,k',k''}    \frac{k'' }{\langle q \rangle} P(q,q',q'',k,k',k'')
\Bigl\{ (1{-}x')^{q'} (1{-}y)^{k} (1{-}y'')^{k''}
\\
+(1{-}x'')^{q''} (1{-}y')^{k'} (1{-}y'')^{k''}
- (1{-}x')^{q'} (1{-}x'')^{q''} (1{-}y)^{k} (1{-}y')^{k'} (1{-}y'')^{k''}
\Bigr\},
\end{multlined}
\label{c120}
\\
1 {-} x'  &= \begin{multlined}[t] 
\sum_{q,q',q'',k,k',k''}    \frac{k'' }{\langle q' \rangle} P(q,q',q'',k,k',k'')
\Bigl\{ (1{-}x)^{q} (1{-}y)^{k} (1{-}y')^{k'}
\\
+(1{-}x'')^{q''} (1{-}y')^{k'} (1{-}y'')^{k''}
- (1{-}x)^{q} (1{-}x'')^{q''} (1{-}y)^{k} (1{-}y')^{k'} (1{-}y'')^{k''}
\Bigr\},
\end{multlined}
\label{c130}
\\
1 {-} x''  &= \begin{multlined}[t]  
\sum_{q,q',q'',k,k',k''}    \frac{k'' }{\langle q'' \rangle} P(q,q',q'',k,k',k'')
\Bigl\{ (1{-}x)^{q} (1{-}y)^{k} (1{-}y')^{k'}
\\
+(1{-}x')^{q'} (1{-}y)^{k} (1{-}y'')^{k''}
- (1{-}x)^{q} (1{-}x')^{q'} (1{-}y)^{k} (1{-}y')^{k'} (1{-}y'')^{k''}
\Bigr\},
\end{multlined}
\label{c140}
\\
1 {-} y  &= \begin{multlined}[t]  
\sum_{q,q',q'',k,k',k''}    \frac{k'' }{\langle k \rangle} P(q,q',q'',k,k',k'')
 (1{-}x'')^{q''} (1{-}y')^{k'} (1{-}y'')^{k''}
\end{multlined},
\label{c150}
\\
1 {-} y'  &= \begin{multlined}[t] 
\sum_{q,q',q'',k,k',k''}    \frac{k'' }{\langle k' \rangle} P(q,q',q'',k,k',k'')
 (1{-}x')^{q'} (1{-}y)^{k} (1{-}y'')^{k''}
,
\end{multlined}
\label{c160}
\\
1 {-} y'' &= \begin{multlined}[t]  
\sum_{q,q',q'',k,k',k''}    \frac{k'' }{\langle k'' \rangle} P(q,q',q'',k,k',k'')
 (1{-}x)^{q} (1{-}y)^{k} (1{-}y')^{k'}
.
\end{multlined}
\label{c170}
\end{align}

The equation for $S$ is then
\begin{multline}
1 {-} S  = \!\!\!\!\!\!\!\!\!
\sum_{q,q',q'',k,k',k''} \!\!\!\!\!\!\!\!\! P(q,q',q'',k,k',k'')\Bigl\{
(1{-}x)^q (1{-}y)^{k} (1{-}y')^{k'}
+(1{-}x')^{q'} (1{-}y)^{k} (1{-}y'')^{k''}
+(1{-}x'')^{q''} (1{-}y')^{k'}(1{-}y'')^{k''}
\\
+ (1{-}y)^{k} (1{-}y')^{k'}(1{-}y'')^{k''} \big[ (1{-}x)^q  (1{-}x')^{q'} (1{-}x'')^{q''} 
-(1{-}x)^q(1{-}x')^{q'}
-(1{-}x)^q (1{-}x'')^{q''}
-(1{-}x')^{q'}(1{-}x'')^{q''} 
\big]
\Bigr\}.
\label{c30}
\end{multline}
\end{widetext}

For simplicity, we consider again the uncorrelated case.
Let $P(q,q',q'',k,k',k'')=P(q)P(q')P(q'')Q(k)Q(k')Q(k'')$, and the generating functions of $P(q)$ and $Q(k)$ be $G(z)$ and $R(z)$, respectively.
From the symmetry of the distributions the probabilities $x=x'=x''$ and $y=y'=y''$.
Then
\begin{align}
1 {-} x  &= 
2 G(1{-}x) R(1{-}y)^2 - G(1{-}x)^2 R(1{-}y)^3
,
\label{c210}
\\%
1 {-} y  &=   G(1{-}x) R(1{-}y)^2
,
\label{c220}
\end{align}
and for $1-S$ we have
\begin{equation}
1 {-} S  = 
3 G(1{-}x) R(1{-}y)^2 
-  R(1{-}y)^3 [ 3G(1{-}x)^2 -G(1{-}x)^3 ]
,
\label{c230}
\end{equation}


\subsection{Phase diagram for \ER layers}

For Poisson distributions $P(q)$ and $Q(k)$ with first moments $\langle q \rangle$ and $\langle k \rangle$, respectively, we have:
\begin{align}
x &= 1- 2e^{-\langle q \rangle x - 2 \langle k \rangle y} + e^{-2\langle q \rangle x - 3 \langle k \rangle y} \equiv F_x(x,y)
, 
\label{c240}
\\%
y &= 1- e^{-\langle q \rangle x - 2 \langle k \rangle y}  \equiv F_y(x,y)
,
\label{c250}
\end{align}
and
\begin{equation}
S=1-3e^{-\langle q \rangle x - 2 \langle k \rangle y} + 3 e^{-2\langle q \rangle x - 3 \langle k \rangle y} - e^{-3\langle q \rangle x - 3 \langle k \rangle y}
.
\label{c260}
\end{equation}

The phase diagram for this case is represented in Fig.~\ref{phase_diags} (c). We observe a line of continuous transitions when the mean overlap degree $\langle k\rangle$ is larger and the non-overlap mean degree $\langle q\rangle$ is smaller.  The growth exponent of the giant component above this continuous transition is $2$, in contrast to the case of triple overlaps, when it was $1$. This exponent $2$ is the same as encountered in pure weak percolation in two layers. 
In contrast, when the density of overlaps is low, we observe a line of discontinuous hybrid transitions, of the same kind as observed in the weak percolation problem without overlaps. 
The two curves meet at a tricritical point, at which the height of the discontinuity reaches zero and the transition becomes continuous.

The phase diagram is obtained by solving the system
\begin{align}
F_x(x,y) &= x,
\label{c270}
\\[3pt]
F_y(x,y) &= y,
\label{c280}
\\[3pt]
\text{det}\!
\left( 
\begin{array}{cc}
\partial_x F_x {-} 1 & \partial_y F_x \ \ \ \ 
\\[3pt]
\partial_x F_y \ \ \ \  & \partial_y F_y {-}1
\end{array}
\right) &= 0
.
\label{c290}
\end{align}
In the phase diagram, the line of continuous transitions can be found by simply setting $x,y=0$ in Eq.~(\ref{c290}), which gives 
\begin{equation}
\langle k \rangle_c^\text{cont} = 1/(2+\langle q \rangle)
.
\label{c291}
\end{equation}

\begin{figure}[t]
\begin{center}
\includegraphics[width=\columnwidth]{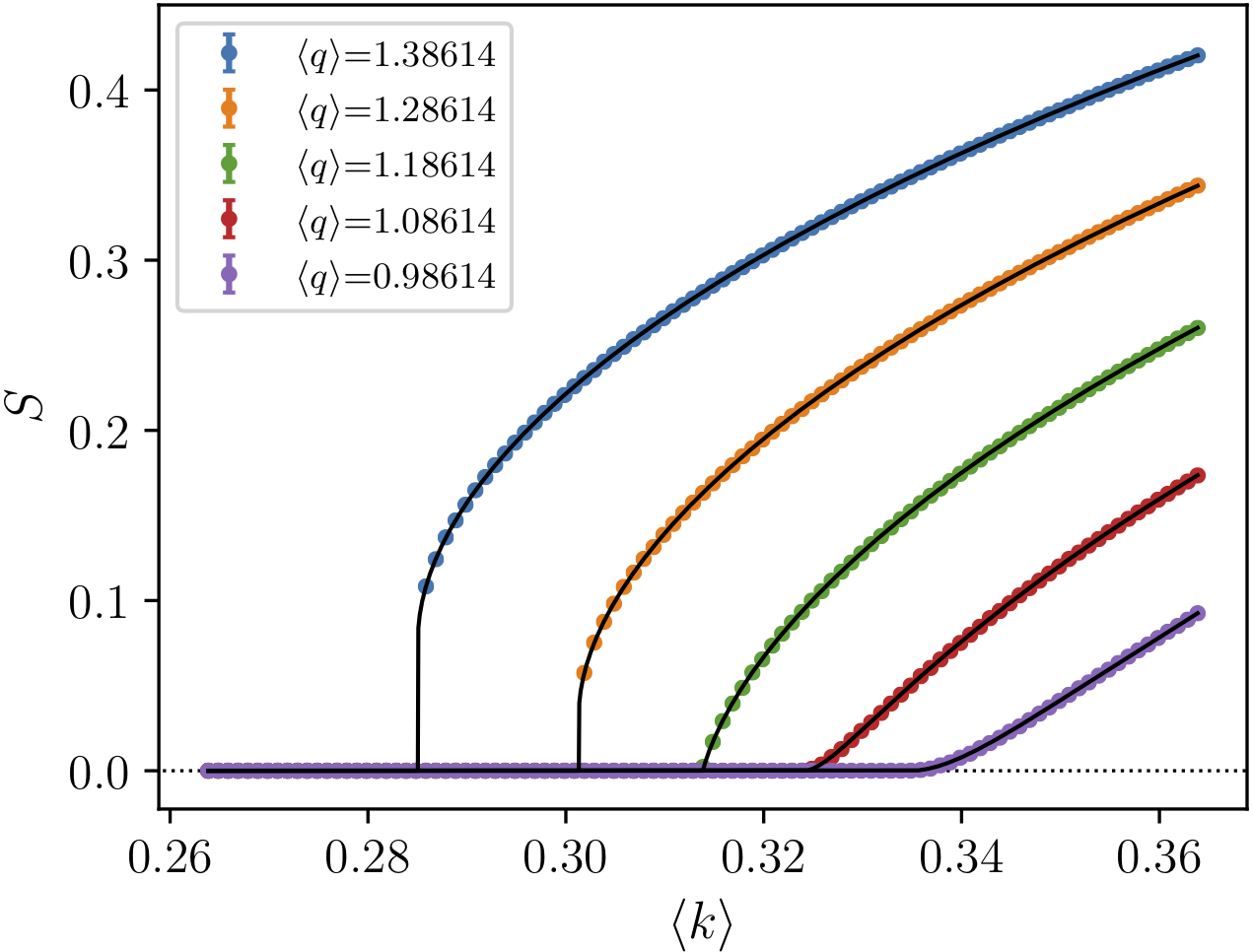} 
\end{center}
\caption{
Simulations of weak percolation in the presence of single and doubly overlapped edges (no triple edges), for symmetrical uncorrelated Poisson degree distributions with means $\langle q \rangle$ and $\langle k \rangle$, respectively.
The middle (green) set of points corresponds to the tricritical point.
Network are generated with $10^8$ nodes, and each point is the average of $10$ realizations.
Black lines represent Eqs.~(\ref{c260}) with $x$ and $y$ replaced by the solution of Eqs.~(\ref{c240}) and (\ref{c250}).
}
\label{f17b}       
\end{figure}

To find the line of discontinuous transitions $\langle k \rangle_c^\text{disc} (\langle q \rangle)$,
let us define the function $\tilde{F}(x) = F_x[x,y^*(x)]$ where $y^*(x)$ is the solution of Eq.~(\ref{c280}) for given $x$.
In terms of the function $\tilde{F}(x)$ the transition point $\langle k \rangle_c  (\langle q \rangle)$ is found by satisfying the two conditions $\tilde{F}(x)=x$ and $\tilde{F}'(x)=1$ simultaneously.
In addition to these two conditions, the tricritical point is determined by the additional condition $\tilde{F}''(x)=0$.
Since the jump in probabilities $x$ and $y$ vanishes at the tricritical point, these three conditions must be met at $x=0$ (which gives also $y^*=0$)
as can be seen in the inset of Fig.~\ref{phase_diags} (c).

The first derivative $\tilde{F}'(x)$ can be written as
\begin{align}
\tilde{F}'(x) =& \frac{\partial F_x(x,y^*)}{\partial x} + \frac{\partial F_x(x,y^*)}{\partial y^*} \frac{\partial y^*}{\partial x}
\nonumber
\\
 \begin{split}
 =&
2\langle q \rangle \left( e^{-\langle q \rangle x - 2 \langle k \rangle y^*} - e^{-2\langle q \rangle x - 3 \langle k \rangle y^*} \right)
\\
&+  \langle q \rangle \langle k \rangle \frac{ 2 e^{-\langle q \rangle x - 2 \langle k \rangle y^*} - 3 e^{-2\langle q \rangle x - 3 \langle k \rangle y^*} }{ e^{\langle q \rangle x + 2 \langle k \rangle y^*} - 2 \langle k \rangle  }\,,
\end{split}\label{c300}
\end{align}
where we replaced $\frac{\partial y^*}{\partial x} = { \langle q \rangle }/{ [e^{\langle q \rangle x + 2 \langle k \rangle y^*} - 2 \langle k \rangle }]$, obtained by differentiating Eq.~(\ref{c250}).
According to Eq.~(\ref{c300}) at $x=0$ (and $y^*=0$) the condition $\tilde{F}'(0)=1$ gives
\begin{equation}
\tilde{F}'(0) = \frac{ \langle q \rangle \langle k \rangle }{ 1- 2 \langle k \rangle  } = 1,
\label{c310}
\end{equation}
which actually describes the line  $\langle k \rangle_c = 1/(2+\langle q \rangle)$ of continuous transitions down to the tricritical point, because along that line the size of the jumps is indeed $0$.

Similarly, the condition for the second derivative becomes
\begin{equation}
\tilde{F}''(0) =  \frac{\langle q \rangle^2 (1-\langle k \rangle) [2-\langle k \rangle(7-2\langle k \rangle)] }{(1- 2 \langle k \rangle)^3} = 0.
\label{c320}
\end{equation}

The tricritical point is found by solving the system formed by Eqs.~(\ref{c310}) and~(\ref{c320}), giving
\begin{align}
\langle q \rangle_\text{T} &= \frac{\sqrt{33}-1}{4} \approx 1.1861...  \,,
\nonumber
\\
 \langle k \rangle_\text{T} &= \frac{7-\sqrt{33}}{4} \approx 0.3138...\,.
\label{c340}
\end{align}

Numerical solutions for $S$ below, above and (almost) at this tricritical point are illustrated in Fig. \ref{f17b}, showing the discontinuous transition above and continuous transition below this point.

\subsection{Behavior near the tricritical point}

Now let us analyse the behavior of the size of the giant component near these transition points, below, above and at the tricritical point.
For small $x$ we expand $\tilde{F}(x)$ up to the cubic term
\begin{equation}
\tilde{F}(x) \cong \tilde{F}(0) + \tilde{F}'(0) x + \frac{\tilde{F}''(0)}{2!} x^2 + \frac{\tilde{F}'''(0)}{3!} x^3 
.
\label{c350}
\end{equation}
We introduce the notation  $\langle q \rangle = \langle q \rangle_\text{T} + \delta$ and $\langle k \rangle_\text{T}+\tilde{\delta}$, where the subscript $\text{T}$ indicates the value at the tricritical point, Eq.~(\ref{c340}). Close to the tricritical point 
the self-consistency equation $\tilde{F}(x)=x$ can then be expressed as
\begin{equation}
(A \delta  {+} \tilde{A} \delta)x + \tilde{B} \tilde{\delta} x^2 + C x^3 = 0,
\label{c380}
\end{equation}
where $A$, $\tilde{A}$, $\tilde{B}$, and $C$ are constants (see 
 \ref{ap_C}).
This equation has two non-trivial solutions (besides $x=0$), and the discontinuous transition occurs when these are equal.

The expansion of Eq.~(\ref{c260}) gives
\begin{equation}
S =  3 \langle k \rangle^2 y^2 + 3 \langle q \rangle \langle k \rangle x y + ... ,
\label{c430}
\end{equation}
which, after first calculating the size of the jumps in $x$ and $y$, gives the size of the jump in $S$ as
\begin{equation}
S_c \cong
\frac{891}{128}(5\sqrt{33}-27){\delta}^2.
\label{c440}
\end{equation}

Note that the slope of the line of the discontinuity at the tricritical point is equal to the slope of the line of continuous transitions at the same point, i.e., $\partial_{\langle q \rangle} \langle k \rangle_c^\text{disc} (\langle q \rangle_\text{T}) = \partial_{\langle q \rangle} \langle k \rangle_c^\text{cont} (\langle q \rangle_\text{T})  = -{(41-7\sqrt{33})}/{8}$.
 

To check the critical exponents of the hybrid transition, let
$ \epsilon =  \langle q \rangle - \langle q \rangle_c$ and $\tilde{\epsilon} = \langle k \rangle - \langle k \rangle_c$ be deviations from a point $ (\langle q \rangle_c, \langle k \rangle_c)$ in the critical line. 
For $ (\langle q \rangle_c, \langle k \rangle_c)$ near the tricritical point, and for small $ \epsilon$ and $ \tilde{\epsilon}$,
near the tricritical point the expansion of $S$ is
\begin{multline}
S \cong S_c + 27 \frac{11{-}\sqrt{33}}{8}  \delta \sqrt{   \frac{3{+}\sqrt{33}}{2}\tilde{ \epsilon} {+} \frac{5 \sqrt{33} {-} 27}{4} \epsilon }
\\
+  3 \frac{3+\sqrt{33}}{4} \left( \frac{3{+}\sqrt{33}}{2} \tilde{\epsilon} {+} \frac{5 \sqrt{33} {-} 27}{4} \epsilon \right)
.
\label{c470a}
\end{multline}
Notice that in Eq.~(\ref{c470a})
 the argument of the square root is positive as long as we remain above the transition line.
Furthermore, the amplitudes of the square-root singular terms 
 for $x$ and $y$, remain finite at the tricritical point, but the one of Eq.~(\ref{c470a}), for $S$, is linear in $ \delta $. 
So, the singularity is square-root away from the tricritical point, but the region where the square-root of $S$ dominates over the linear contributions vanishes approaching the tricritcal point, and, at the tricritical point, the singularity of $S$ has exponent $1$.


Solving $x=\tilde{F}(0) + \tilde{F}'(0) x+ \tilde{F}''(0) x^2/2 $ gives
\begin{equation}
S =  
 \frac{ 12(1 + {\langle q \rangle_c}) }{ {(1 {+} {\langle q \rangle_c} )}^2 {[4 {-} \langle q \rangle_c {-} 2 {\langle q \rangle_c}^2] }^2 }  {[ \epsilon + (2 {+} \langle q \rangle_c)^2  \tilde{\epsilon} ]}^2
 .
\label{c510a}
\end{equation}

Thus, over the line of continuous transitions the singularity has exponent $2$, as expected, since $x$ and $y$ grow linearly, and $S \propto x^2$. 

The exponent exactly at the tricritical point, however, is $1$.
As shown above, at that special point we must consider terms up to the third derivative $\tilde{F}'''(0)$.
So, exactly at the tricritical point Eq.~(\ref{c510a}) does not apply, but we can see from  Eq.~(\ref{c470a}), that at the tricritical point the exponent should be $1$.


\section{Discussions and conclusions}
\label{s5} 

In this paper we have studied the phase diagrams and critical phenomena of the emergence of the giant weak percolation component in the presence of overlapping edges, in two- and three-layer multiplex networks.
We found that the presence of overlaps  can alter the universality class of the transition. In two layers, the transition without overlaps has $\beta$-exponent $2$, but with any amount of overlapping edges this reduces to $1$, as found in ordinary percolation. Meanwhile, in three layers, the transition in the absence of overlaps is a discontinuous hybrid transition, of the $k$-core type, but the presence of overlaps in sufficient density leads instead to a continuous transition. Furthermore, if the overlaps are only between two layers at a time, we find a tricritical point where the lines of discontinuous and continuous transitions meet.

This behavior contrasts with the alternative percolation process in multi-layer networks, the emerge of the giant mutually connected component. In that problem, overlaps do not alter the nature of the transition. In particular, the discontinuous transition in two or more layers is not destroyed with any amount of overlaps \cite{hu2013percolation, min2015link, cellai2016message, baxter2016correlated}.
If the overlaps are in all layers of the multiplex (double overlaps in two layers, or triple overlaps in three layers) we may, following Refs.~\cite{hu2013percolation, min2015link}, treat a multiplex network with overlapping edges as a set of super-nodes---each one representing a cluster connected only by overlaps---interconnected by single edges belonging to individual layers. 
In the interdependency problem, 
the super-nodes can be reduced to single vertices without changing the global connectivity of the network. 
Thus this problem with overlaps can be always reduced, in essence, to the problem without them, and consequently even a small concentration of single edges determines the nature of the transition. 
In contrast to this, in the weak percolation problem, a single edge between two super-nodes effectively merges them together 
and so we arrive at a situation similar to ordinary percolation.

In three layers, the phase diagram, Fig.  \ref{phase_diags}~(c), for only single edges and double overlaps, is qualitatively similar to that observed in the $(2,3)$-heterogeneous $k$-core, where  a tricritical point was also observed \cite{cellai2011tricritical}. Both problems contain a mixture of vertices requiring either two or three connections in order to belong to the giant component. 
On the other hand, the three-layer weak percolation problem with only triple overlaps and single edges produces a phase diagram, Fig. \ref{phase_diags}~(b), similar to that observed in the $(1,3)$-heterogeneous $k$-core \cite{baxter2011heterogeneous}. In both cases, vertices require either one or three connections of the right kind to belong to the giant component.
We also expect that the dynamics near the discontinuous transitions should be similar to that for the $k$-core problem \cite{baxter2015critical}.

The techniques we have developed allow for the straightforward derivation of  equations for this class of problems, and while more onerous, can easily be generalised to higher numbers of layers. We suggest that the phase diagrams for four or higher number of layers will be similar to what we observed for three layers.

\section*{Acknowledgement}

This work was developed within the scope of the project i3N, UIDB/50025/2020 and UIDP/50025/2020, financed by national funds through the FCT/MEC. This work was also supported by National Funds through FCT, I. P. Project No. IF/00726/2015. R. A. d. C. acknowledges the FCT Grants No. SFRH/BPD/123077/2016 and No. CEECIND/04697/2017.

\appendix


\section{Derivation of equations for two layers}
\label{sa1} 

Here we strictly derive Eqs.~(\ref{1196})--(\ref{1210}) by listing all contributions to their right-hand sides.

In our previous works, we used a graphical technique for derivations of similar equations, where each involved probability is shown by an edge in a given layer, leading to a finite or infinite component, e.g., see Fig. \ref{f13}. 
In this technique, each contribution to the right-hand side of self-consistency  equations for involved probabilities and of an expression for the relative size of a giant component is represented in a graphical form. This approach enables us to conveniently list all contributions and then write out a polynomial of relevant probabilities for each of them. 
For the problems considered in this work, the graphical form turns out to be too cumbersome, and it is more convenient to show each contribution as a list of numbers of different edges  leading to finite or infinite components. For this list we easily write a polynomial of probabilities. 

In the weak percolation problem with overlaps, we introduce five probabilities, and, hence, five types of edges in accordance to where these edges lead, see Fig.~\ref{f13}. 
The sixth probability, $w$, that a double edge leads to a finite weak component equals $1-z$. 
Note that for $k$ (degree) double edges, the probability that they lead {\em only} to finite weak components equals $(1-\delta_{k,0})(1-z)^k$. Here $1-\delta_{k,0}$ is necessary since this probability equals zero if $k=0$ and not $1$. This $1-\delta_{k,0}$ demands a separate consideration of $k=0$ and $k\geq1$, which slightly complicates our analysis.

For each of five kinds of edges, it is sufficient to consider two sets of their numbers: $0$ and $\geq 1$. This gives $2^5$ combinations in total. In fact the number of relevant combinations is much less. The union of $0$ and $\geq 1$ gives $\geq 0$ or ``any'' resulting in the probability equal to $1$. 

($S$) Let us begin with the equation for $S$.

As preparation we show all relevant combinations of $q$ balls of two colors: the white balls (probability $x$) and the black balls (probability $u$): 
\begin{align}
&x  &u   & &   &\text{probability}
\nonumber
\\[5pt]
\geq &0  &\geq 0  & &\longrightarrow \qquad & 1
\nonumber
\\[3pt]
\geq &1   &\geq 0 & &\longrightarrow  \qquad & 1 - (1-x)^q
\nonumber
\\[3pt]
&0 &\geq 0 & &\longrightarrow \qquad &(1-x)^q 
\nonumber
\\[3pt]
\geq &1 &  \geq 1 & &\longrightarrow \qquad &1 - (1{-}x)^q - (1{-}u)^q + (1{-}x{-}u)^q
\nonumber
\\[3pt]
&0 &  \geq 1& &\longrightarrow \qquad &(1-x)^q - (1-x-u)^q 
\nonumber
\\[3pt]
& 0 & 0& &\longrightarrow \qquad &(1-x-u)^q 
.
\label{a20}
\end{align}
One can check the three last lines. It must be that
\begin{equation}
(0,0) + (0,\geq 1) + (\geq 1,0) + (\geq 1,\geq 1) = (\geq 0,\geq 0) = 1 
,
\label{a30}
\end{equation}
where $(x,y)$ denotes the corresponding probability, which is indeed the case.

\begin{figure}[h]
\begin{center}
\includegraphics[width=0.8\columnwidth]{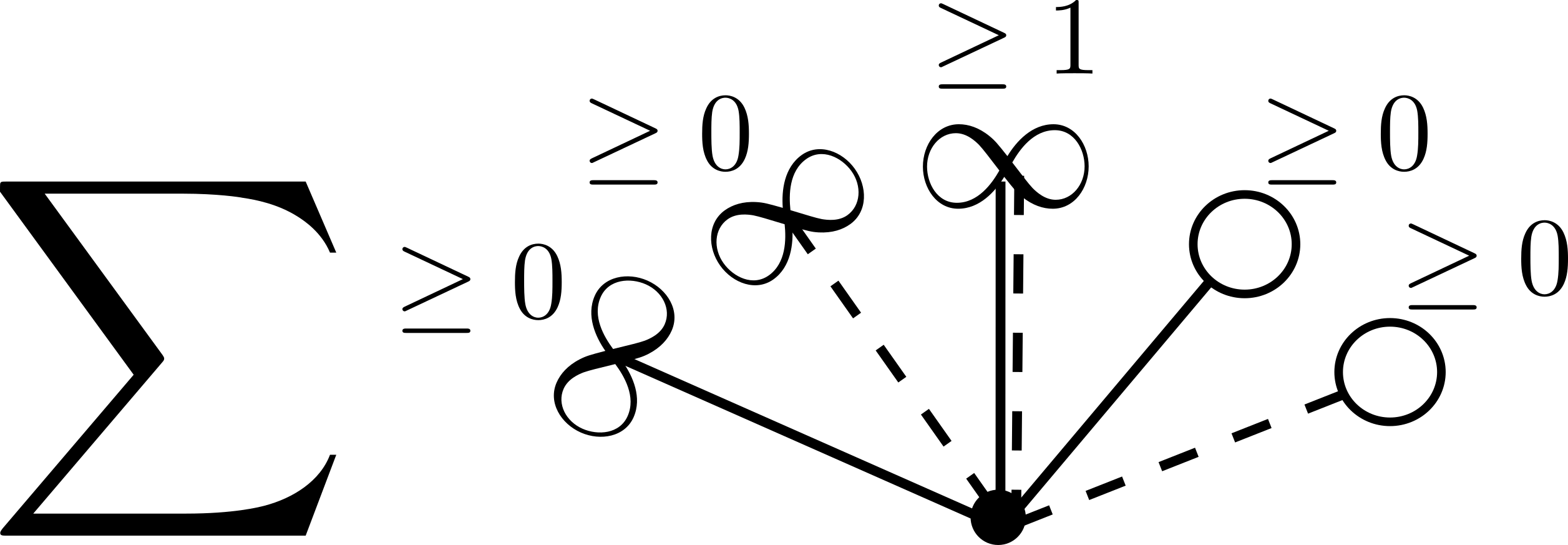} 
\end{center}
\caption{
Graphical representation of the first line of Eq. (\ref{a10}) listing contributions to $S$. See the notations in Fig. \ref{f13}.
}
\label{S_derivation}       
\end{figure}

For $S$ we have  the following set of relevant combinations (lists of length $5$):

\begin{eqnarray}
&& 
\phantom{1-\delta_{k,0} \ \ \ \ \ \ \ \  } x \ \ \ \ \ \ \ y \ \ \ \ \ \ \ z \ \ \ \ \ \ \ u \ \ \ \ \ \ \ v   
\nonumber
\\[6pt]
&& 
\phantom{1-\delta_{k,0} \ \ \ \ \ \ \ \  } \!\!{\geq}0 \ \ \ \ \  {\geq}0\ \ \ \ \, {\geq}1\ \ \ \ \  {\geq}0 \ \ \ \ \,  {\geq}0 
\nonumber
\\[6pt]
&& 
\phantom{1-\delta_{k,0} \ \ \ \ \ \ \ \  } \!\!{\geq}1 \ \ \ \ \ \  \!\!{\geq}1 \ \ \ \ \ \, 0 \ \ \ \ \ \ {\geq}0 \ \ \ \ \  {\geq}0  
\nonumber
\\[2pt]
&& 
1-\delta_{k,0} \ \ \ \ \ \ \ \   \!\!{\geq}1 \ \ \ \ \ \ 0\ \ \ \ \ \ \ 0 \ \ \ \ \ \ {\geq}0 \ \ \ \ \  {\geq}0  
\nonumber
\\[2pt]
&& 
\phantom{1-}\ \, \delta_{k,0} \ \ \ \ \ \ \ \   \!\!{\geq}1 \ \ \ \ \ \ 0\ \ \ \ \ \ \ 0 \ \ \ \ \ \ {\geq}0 \ \ \ \ \  {\geq}1 
\nonumber
\\[6pt]
&& 
1-\delta_{k,0} \ \ \ \ \ \ \ \   0 \ \ \ \ \ \ {\geq}1\ \ \ \ \ \ 0 \ \ \ \ \ \ {\geq}0 \ \ \ \ \  {\geq}0  
\nonumber
\\[2pt]
&& 
\phantom{1-}\ \, \delta_{k,0} \ \ \ \ \ \ \ \   0 \ \ \ \ \ \ {\geq}1\ \ \ \ \ \ 0 \ \ \ \ \ \ {\geq}1 \ \ \ \ \ {\geq}0  
\label{a10}
\end{eqnarray}
The combination corresponding to the first line is illustrated in Fig.~\ref{S_derivation}.
Note that for $k$ (degree) double edges, the probability that they lead {\em only} to finite weak components equals $(1-\delta_{k,0})(1-z)^k$. Here $1-\delta_{k,0}$ is necessary since this probability equals zero if $k=0$ and not $1$. This factor $(1-\delta_{k,0})$, 
which is absent in the theory of interdependent network with overlaps,   
demands a separate consideration of $k=0$ and $k\geq1$ and plays an important role in our analysis.
In the left column we specially indicate the combinations where the number of overlaps is not $0$ (factor $1-\delta_{k,0}$) or where this number is $0$ (factor $\delta_{k,0}$). 

Equation~(\ref{a10}) corresponds to 
\begin{multline}
S
= 
\sum_{q,q',k} P(q,q',k)\{ 
1\cdot 1\cdot [1 - (1{-}z)^k] 
\\[3pt]
+ [1 {-} (1{-}x)^q] [1 {-} (1{-}y)^{q'}] (1{-}z)^k 
\\[3pt]
+ (1-\delta_{k,0})[1 {-} (1{-}x)^q] (1{-}y)^{q'} (1{-}z)^k 
\\[3pt]
+  \delta_{k,0}[1 {-} (1{-}x)^q] [(1{-}y)^{q'} {-} (1{-}y{-}v)^{q'}]  (1{-}z)^k
\\[3pt]
+  (1-\delta_{k,0})(1{-}x)^q [1 {-} (1{-}y)^{q'}] (1{-}z)^k
\\[3pt]
+  \delta_{k,0}[(1{-}x)^q {-} (1{-}x{-}u)^q] [1 {-} (1{-}y)^{q'}] (1{-}z)^k\}
, 
\label{a40}
\end{multline}
which can be reduced to Eq. (\ref{1210}).

There are five equations for $x$, $y$, $z$, $u$, $v$.

($x$) The right-hand side of the self-consistency expression for $x$ has the following contributions:
\begin{eqnarray}
&& 
\phantom{1-\delta_{k,0} \ \ \ \ \ \ \ \  } x \ \ \ \ \ \ \ y \ \ \ \ \ \ \ z \ \ \ \ \ \ \ u \ \ \ \ \ \ \ v   
\nonumber
\\[6pt]
&& 
\phantom{1-\delta_{k,0} \ \ \ \ \ \ \ \  } \!\!{\geq}0 \ \ \ \ \  {\geq}0\ \ \ \ \, {\geq}1\ \ \ \ \  {\geq}0 \ \ \ \ \,  {\geq}0  
\nonumber
\\[6pt]
&& 
\phantom{1-\delta_{k,0} \ \ \ \ \ \ \ \  } \!\!{\geq}0 \ \ \ \ \,  {\geq}1\ \ \ \ \ \ 0 \ \ \ \ \ \ {\geq}0 \ \ \ \ \ {\geq}0  
\nonumber
\\[6pt]
&& 
1-\delta_{k,0} \ \ \ \ \ \ \ \   \!\!{\geq}1 \ \ \ \ \ \ 0 \ \ \ \ \ \ \ 0 \ \ \ \ \ \  {\geq}0 \ \ \ \ \,  {\geq}0  
\nonumber
\\[6pt]
&& 
\phantom{1-}\ \, \delta_{k,0}  \ \ \ \ \ \ \   {\geq}1 \ \ \ \ \ \ 0 \ \ \ \ \ \ \ 0 \ \ \ \ \ \ {\geq}0 \ \ \ \ \, {\geq}1 
\label{a60}
\end{eqnarray}
which corresponds to
\begin{multline}
x =  
\sum_{q,q',k} \frac{q}{\langle q \rangle}P(q,q',k)\bigl\{ 
1\cdot1\cdot [1 {-} (1{-}z)^k]
\\[3pt]
+ 1\cdot [1 {-} (1{-}y)^{q'}] (1{-}z)^k
\\[3pt]
+
(1-\delta_{k,0})[1 - (1{-}x)^{q-1}] (1{-}y)^{q'}(1{-}z)^k 
\\[3pt]
+
\delta_{k,0}[1 - (1{-}x)^{q-1}] [(1{-}y)^{q'} - (1{-}y{-}v)^{q'}] 
\bigr\} 
\label{a70}
\end{multline}
and which can be reduced to Eq.~(\ref{1196}). 

($y$) The expression for $y$ contains the combinations 
\begin{eqnarray}
&& 
\phantom{1-\delta_{k,0} \ \ \ \ \ \ \ \  } x \ \ \ \ \ \ \ y \ \ \ \ \ \ \ z \ \ \ \ \ \ \ u \ \ \ \ \ \ \ v   
\nonumber
\\[6pt]
&& 
\phantom{1-\delta_{k,0} \ \ \ \ \ \ \ \  } \!\!{\geq}0 \ \ \ \ \  {\geq}0\ \ \ \ \, {\geq}1\ \ \ \ \  {\geq}0 \ \ \ \ \,  {\geq}0  
\nonumber
\\[6pt]
&& 
\phantom{1-\delta_{k,0} \ \ \ \ \ \ \ \  } \!\!{\geq}1 \ \ \ \ \   {\geq}0\ \ \ \ \ \  0 \ \ \ \ \ \ {\geq}0 \ \ \ \ \, {\geq}0 
\nonumber
\\[6pt]
&& 
1-\delta_{k,0} \ \ \ \ \ \ \ \ \, 0 \ \ \ \ \ \  {\geq}1\ \ \ \ \ \ 0 \ \ \ \ \ \  {\geq}0 \ \ \ \ \,  {\geq}0  
\nonumber
\\[6pt]
&& 
\phantom{1-}\ \, \delta_{k,0}  \ \ \ \ \ \ \ \ \, 0 \ \ \ \ \ \   {\geq}1\ \ \ \ \ \   0 \ \ \ \ \ \ {\geq}1  \ \ \ \ \, {\geq}0 
\label{a80}
\end{eqnarray}
which corresponds to 
\begin{multline}
y 
=
\sum_{q,q',k} \frac{q'}{\langle q' \rangle}P(q,q',k)\bigl\{ 
1\cdot1\cdot [1 - (1{-}z)^k]
\\[3pt]
+
1\cdot [1 - (1{-}x)^q] (1{-}z)^k
\\[3pt]
+
(1-\delta_{k,0}) (1{-}x)^q [1 - (1{-}y)^{q'-1}]  (1{-}z)^k
\\[3pt]
+
\delta_{k,0} [ (1{-}x)^q - (1{-}x{-}u)^q ] [1 - (1{-}y)^{q'-1}]
\bigr\}
\label{a90}
\end{multline}
leading to Eq.~(\ref{1197}).

($z$) The right-hand side of the expression for $z$ has the contributions
\begin{eqnarray}
&& 
x \ \ \ \ \ \ \ y \ \ \ \ \ \ \ z \ \ \ \ \ \ \ u \ \ \ \ \ \ \ v   
\nonumber
\\[6pt]
&& 
\!\!{\geq}0 \ \ \ \ \  {\geq}0\ \ \ \ \, {\geq}1\ \ \ \ \  {\geq}0 \ \ \ \ \,  {\geq}0  
\nonumber
\\[2pt]
&& 
\!\!{\geq}1 \ \ \ \ \ \ 0\ \ \ \ \ \ \ 0 \ \ \ \ \ \ {\geq}0 \ \ \ \ \,  {\geq}0  
\nonumber
\\[2pt]
&& 
0 \ \ \ \ \ \ {\geq}1\ \ \ \ \ \ 0 \ \ \ \ \ \ {\geq}0 \ \ \ \ \, {\geq}0 
\nonumber
\\[6pt]
&& 
\!\!{\geq}1 \ \ \ \ \,  {\geq}1\ \ \ \ \ \ 0 \ \ \ \ \ \ {\geq}0 \ \ \ \ \, {\geq}0 
\label{a100}
\end{eqnarray}
which corresponds to 
\begin{multline}
z =
\sum_{q,q',k} \frac{k}{\langle k \rangle}P(q,q',k)\{ 
1\cdot1\cdot [1 {-} (1{-}z)^{k-1}]
\\[3pt]
+ [1 {-} (1{-}x)^q] (1{-}y)^{q'} (1{-}z)^{k-1}
\\[3pt]
+ (1{-}x)^q [1 {-} (1{-}y)^{q'}] (1{-}z)^{k-1}
\\[3pt]
+ [1 {-} (1{-}x)^q] [1 {-} (1{-}y)^{q'}] (1{-}z)^{k-1}
\}
\label{a110}
\end{multline}
resulting in Eq.~(\ref{1198}).

($u$) The expression for $u$ is:
\begin{eqnarray}
&& 
\phantom{1-\delta_{k,0} \ \ \ \ \ \ \ \ } x \ \ \ \ \ \ \ y \ \ \ \ \ \ \ z \ \ \ \ \ \ \ u \ \ \ \ \ \ \ v   
\nonumber
\\[6pt]
&&  
1-\delta_{k,0} \ \ \ \ \ \ \ \ 0 \ \ \ \ \ \ \  0\ \ \ \ \ \ \ 0 \ \ \ \ \ \ {\geq}0 \ \ \ \ \  {\geq}0  
\nonumber
\\[2pt]
&& 
\phantom{1-}\delta_{k,0} \ \ \ \ \ \ \ \ \ \,  0 \ \ \ \ \ \ \ 0\ \ \ \ \ \ \ 0 \ \ \ \ \ \ {\geq}0 \ \ \ \ \ {\geq}1  
\label{a120}
\end{eqnarray}
which corresponds to 
\begin{multline}
u =
\sum_{q,q',k} \frac{q}{\langle q \rangle}P(q,q',k)\{ 
(1-\delta_{k,0}) (1{-}x)^{q-1} (1{-}y)^{q'}(1{-}z)^k
\\[3pt]
+ \delta_{k,0} (1{-}x)^{q-1} [(1{-}y)^{q'} {-} (1{-}y{-}v)^{q'}]
. 
\label{a130}
\end{multline}
Eqs. (\ref{a70}) and (\ref{a130}) then lead to Eq.~(\ref{1199}).

($v$) Finally, the expression for $v$ contains
\begin{eqnarray}
&& 
\phantom{1-\delta_{k,0} \ \ \ \ \ \ \ \ } x \ \ \ \ \ \ \ y \ \ \ \ \ \ \ z \ \ \ \ \ \ \ u \ \ \ \ \ \ \ v  
\nonumber
\\[6pt]
&& 
1-\delta_{k,0} \ \ \ \ \ \ \ \  0 \ \ \ \ \ \ \ \, 0\ \ \ \ \ \ \ 0 \ \ \ \ \ \ {\geq}0 \ \ \ \ \  {\geq}0  
\nonumber
\\[2pt]
&& 
\phantom{1-}\delta_{k,0} \ \ \ \ \ \ \ \ \ \, 0 \ \ \ \ \ \ \ \, 0\ \ \ \ \ \ \ 0 \ \ \ \ \ \  {\geq}1 \ \ \ \ \  {\geq}0  
\label{a140}
\end{eqnarray}
which corresponds to 
\begin{multline}
v =
\sum_{q,q',k} \frac{q'}{\langle q' \rangle}P(q,q',k)\{ 
(1-\delta_{k,0}) (1{-}x)^q (1{-}y)^{q'-1} (1{-}z)^k
\\[3pt]
+\delta_{k,0} [(1{-}x)^q {-} (1{-}x{-}u)^q] (1{-}y)^{q'-1}
 \}
.
\label{a150}
\end{multline}
Eqs. (\ref{a90}) and (\ref{a150}) lead to Eq.~(\ref{1200}).

\section{Equations for three layers with triple overlaps}
\label{sa33}

Similarly to Fig.~\ref{f13}, we introduce the probabilities $x$, $y$, $z$, $r$, $u$, $v$, $w$, $s = 1-r$, Fig.~\ref{f5}, for which we should write seven equations and an expression for the relative size $S$ of the giant component. These equations and the expression can be derived strictly by following the derivation in 
 \ref{sa1} and Eqs.~(\ref{1196})--(\ref{1210}).  

\begin{widetext}
($S$) For $S$ we have  
\begin{eqnarray}
&& 
\phantom{1-\delta_{k,0} \ \ \ \ \ \ \ \  } x \ \ \ \ \ \ \ y \ \ \ \ \ \ \ z \ \ \ \ \ \ \ r \ \ \ \ \ \ \ u \ \ \ \ \ \ \ v \ \ \ \ \ \ \ w     
\nonumber
\\[6pt]
&& 
\phantom{1-\delta_{k,0} \ \ \ \ \ \ \ \  } \!\!{\geq}0 \ \ \ \ \  {\geq}0\ \ \ \ \, {\geq}0\ \ \ \ \  {\geq}1 \ \ \ \ \,  {\geq}0 \ \ \ \ \,  {\geq}0 \ \ \ \ \   {\geq}0 
\nonumber
\\[3pt]
&& 
\phantom{1-\delta_{k,0} \ \ \ \ \ \ \ \  } \!\!{\geq}1 \ \ \ \ \ {\geq}1 \ \ \ \ \ \  \!\!{\geq}1 \ \ \ \ \ \, 0 \ \ \ \ \ \ {\geq}0 \ \ \ \ \,  {\geq}0   \ \ \ \ \  {\geq}0
\nonumber
\\[2pt]
&& 
1-\delta_{k,0} \ \ \ \ \ \ \ \   \!\!{\geq}1 \ \ \ \ \  {\geq}1 \ \ \ \ \ \ 0\ \ \ \ \ \ \ 0 \ \ \ \ \ \ {\geq}0 \ \ \ \ \,  {\geq}0  \ \ \ \ \  {\geq}0   
\nonumber
\\[2pt]
&& 
1-\delta_{k,0} \ \ \ \ \ \ \ \   \!\!{\geq}1 \ \ \ \ \ \ 0 \ \ \ \ \ \ {\geq}1 \ \ \ \ \ \ 0 \ \ \ \ \ \ {\geq}0 \ \ \ \ \,  {\geq}0  \ \ \ \ \  {\geq}0   
\nonumber
\\[2pt]
&& 
1-\delta_{k,0} \ \ \ \ \ \ \ \   0 \ \ \ \ \ \ {\geq}1 \ \ \ \ \  {\geq}1 \ \ \ \ \ \ 0 \ \ \ \ \ \ {\geq}0 \ \ \ \ \,  {\geq}0  \ \ \ \ \  {\geq}0   
\nonumber
\\[2pt]
&& 
1-\delta_{k,0} \ \ \ \ \ \ \ \   \!\!{\geq}1 \ \ \ \ \ \ 0 \ \ \ \ \ \ \ \, 0\ \ \ \ \ \ \  0 \ \ \ \ \ \ {\geq}0 \ \ \ \ \,  {\geq}0  \ \ \ \ \  {\geq}0   
\nonumber
\\[2pt]
&& 
1-\delta_{k,0} \ \ \ \ \ \ \ \   0 \ \ \ \ \ \  {\geq}1 \ \ \ \ \ \ \, 0\ \ \ \ \ \ \ 0 \ \ \ \ \ \ {\geq}0 \ \ \ \ \,  {\geq}0  \ \ \ \ \  {\geq}0   
\nonumber
\\[2pt]
&& 
1-\delta_{k,0} \ \ \ \ \ \ \ \   0 \ \ \ \ \ \ \  0 \ \ \ \ \ \ \, {\geq}1 \ \ \ \ \ \ 0 \ \ \ \ \ \ {\geq}0 \ \ \ \ \,  {\geq}0  \ \ \ \ \  {\geq}0   
\nonumber
\\[2pt]
&& 
\phantom{1-}\ \, \delta_{k,0} \ \ \ \ \ \ \ \   \!\!{\geq}1 \ \ \ \ \  {\geq}1 \ \ \ \ \ \ 0\ \ \ \ \ \ \ 0 \ \ \ \ \ \ {\geq}0 \ \ \ \ \,  {\geq}0  \ \ \ \ \  {\geq}1 
\nonumber
\\[2pt]
&& 
\phantom{1-}\ \, \delta_{k,0} \ \ \ \ \ \ \ \   \!\!{\geq}1 \ \ \ \ \ \,  0 \ \ \ \ \ \ \, {\geq}1 \ \ \ \ \ \  0 \ \ \ \ \ \ {\geq}0 \ \ \ \ \,  {\geq}1  \ \ \ \ \  {\geq}0    
\nonumber
\\[2pt]
&& 
\phantom{1-}\ \, \delta_{k,0} \ \ \ \ \ \ \ \   0 \ \ \ \ \ \  {\geq}1 \ \ \ \ \  {\geq}1 \ \ \ \ \ \  0 \ \ \ \ \ \ {\geq}1 \ \ \ \ \,  {\geq}0  \ \ \ \ \  {\geq}0  
\nonumber
\\[2pt]
&& 
\phantom{1-}\ \, \delta_{k,0} \ \ \ \ \ \ \ \   \!\!{\geq}1 \ \ \ \ \ \, 0 \ \ \ \ \ \ \ \, 0\ \ \ \ \ \ \ \, 0 \ \ \ \ \ \ {\geq}0 \ \ \ \ \,  {\geq}1  \ \ \ \ \  {\geq}1  
\nonumber
\\[2pt]
&& 
\phantom{1-}\ \, \delta_{k,0} \ \ \ \ \ \ \ \   0 \ \ \ \ \ \  {\geq}1 \ \ \ \ \ \ 0\ \ \ \ \ \ \ \, 0 \ \ \ \ \ \ {\geq}1 \ \ \ \ \,  {\geq}0  \ \ \ \ \  {\geq}1  
\nonumber
\\[2pt]
&& 
\phantom{1-}\ \, \delta_{k,0} \ \ \ \ \ \ \ \   0 \ \ \ \ \ \ \  0 \ \ \ \ \ \ {\geq}1\ \ \ \ \ \ \, 0 \ \ \ \ \ \ {\geq}1 \ \ \ \ \,  {\geq}1  \ \ \ \ \  {\geq}0 
 \nonumber \\ \label{b9}
\end{eqnarray}

Hence the expression for $1-S$ has the form:  
\begin{multline}
1 - S = 
\sum_{q,q',q'',k} P(q,q',q'',k)\Bigl\{
[1 -  \delta_{k,0}] (1{-}x)^q (1{-}y)^{q'} (1{-}z)^{q''} (1{-}r)^k 
+ \delta_{k,0}\bigl\{ 
(1{-}x)^q (1{-}y)^{q'} (1{-}z)^{q''} 
\\
+ 
[1 {-} (1{-}x)^q] [(1{-}y{-}v)^{q'} (1{-}z)^{q''} + (1{-}y)^{q'} (1{-}z{-}w)^{q''} - (1{-}y{-}v)^{q'} (1{-}z{-}w)^{q''} ] 
\\
+  
[1 {-} (1{-}y)^{q'}] 
[ (1{-}x{-}u)^q (1{-}z)^{q''} + (1{-}x)^q (1{-}z{-}w)^{q''} - (1{-}x{-}u)^q(1{-}z{-}w)^{q''} ]
\\
+ 
[1 {-} (1{-}z)^{q''}] [ (1{-}x{-}u)^q (1{-}y)^{q'} + (1{-}x)^q (1{-}y{-}v)^{q'} - (1{-}x{-}u)^q (1{-}y{-}v)^{q'} ] 
\\
+ 
[1 {-} (1{-}x)^q] [1 {-} (1{-}y)^{q'}] (1{-}z{-}w)^{q''} \!
+  
[1 {-} (1{-}x)^q] (1{-}y{-}v)^{q'} [1 {-} (1{-}z)^{q''}] 
+ 
(1{-}x{-}u)^q [1 {-} (1{-}y)^{q'}] [1 {-} (1{-}z)^{q''}]
\bigr\} \!\!
\Bigr\}
\label{b10}
\end{multline}
which can be written more compactly as Eq. (\ref{b20}).
\end{widetext}

($x$) For $x$ we have  
\begin{eqnarray}
&& 
\phantom{1-\delta_{k,0} \ \ \ \ \ \ \ \  } x \ \ \ \ \ \ \ y \ \ \ \ \ \ \ z \ \ \ \ \ \ \ r \ \ \ \ \ \ \ u \ \ \ \ \ \ \ v \ \ \ \ \ \ \ w     
\nonumber
\\[6pt]
&& 
\phantom{1-\delta_{k,0} \ \ \ \ \ \ \ \  } \!\!{\geq}0 \ \ \ \ \  {\geq}0\ \ \ \ \, {\geq}0\ \ \ \ \  {\geq}1 \ \ \ \ \,  {\geq}0 \ \ \ \ \,  {\geq}0 \ \ \ \ \   {\geq}0 
\nonumber
\\[3pt]
&& 
\phantom{1-\delta_{k,0} \ \ \ \ \ \ \ \  } 0 \ \ \ \ \ \ {\geq}1 \ \ \ \ \ \  \!\!{\geq}1 \ \ \ \ \ \, 0 \ \ \ \ \ \ {\geq}0 \ \ \ \ \,  {\geq}0   \ \ \ \ \  {\geq}0
\nonumber
\\[2pt]
&& 
1-\delta_{k,0} \ \ \ \ \ \ \ \   \!\!{\geq}0 \ \ \ \ \ \  0 \ \ \ \ \ \ {\geq}1 \ \ \ \ \ \, 0 \ \ \ \ \ \ {\geq}0 \ \ \ \ \,  {\geq}0  \ \ \ \ \  {\geq}0   
\nonumber
\\[2pt]
&& 
1-\delta_{k,0} \ \ \ \ \ \ \ \   \!\!{\geq}0 \ \ \ \ \  {\geq}1 \ \ \ \ \ \ 0 \ \ \ \ \ \ \, 0 \ \ \ \ \ \ {\geq}0 \ \ \ \ \,  {\geq}0  \ \ \ \ \  {\geq}0   
\nonumber
\\[2pt]
&& 
1-\delta_{k,0} \ \ \ \ \ \ \ \   \!\!{\geq}1 \ \ \ \ \ \ 0 \ \ \ \ \ \ \ \, 0\ \ \ \ \ \ \,  0 \ \ \ \ \ \ {\geq}0 \ \ \ \ \,  {\geq}1  \ \ \ \ \  {\geq}0   
\nonumber
\\[2pt]
&& 
\phantom{1-}\ \, \delta_{k,0} \ \ \ \ \ \ \ \   \!\!{\geq}0 \ \ \ \ \ \  0 \ \ \ \ \ \, {\geq}1 \ \ \ \ \ \  0 \ \ \ \ \ \ {\geq}0 \ \ \ \ \,  {\geq}1  \ \ \ \ \  {\geq}0    
\nonumber
\\[2pt]
&& 
\phantom{1-}\ \, \delta_{k,0} \ \ \ \ \ \ \ \   \!\!{\geq}0 \ \ \ \ \   {\geq}1 \ \ \ \ \ \  0 \ \ \ \ \ \ \,  0 \ \ \ \ \ \ {\geq}0 \ \ \ \ \,  {\geq}0  \ \ \ \ \  {\geq}1  
\nonumber
\\[2pt]
&& 
\phantom{1-}\ \, \delta_{k,0} \ \ \ \ \ \ \ \   \!\!{\geq}1 \ \ \ \ \ \, 0 \ \ \ \ \ \ \ \, 0\ \ \ \ \ \ \ \, 0 \ \ \ \ \ \ {\geq}0 \ \ \ \ \,  {\geq}1  \ \ \ \ \  {\geq}1  
\nonumber \\
\label{b22}
\end{eqnarray}
and one can write similar arrays for $y$ and $z$.

($r$) For $r$ we have  
\begin{eqnarray}
&& 
\phantom{1-\delta_{k,0} \ \ \ \ \ \ \ \  } x \ \ \ \ \ \ \ y \ \ \ \ \ \ \ z \ \ \ \ \ \ \ r \ \ \ \ \ \ \ u \ \ \ \ \ \ \ v \ \ \ \ \ \ \ w     
\nonumber
\\[6pt]
&& 
\phantom{1-\delta_{k,0} \ \ \ \ \ \ \ \  } \!\!{\geq}0 \ \ \ \ \  {\geq}0\ \ \ \ \, {\geq}0\ \ \ \ \  {\geq}1 \ \ \ \ \,  {\geq}0 \ \ \ \ \,  {\geq}0 \ \ \ \ \   {\geq}0 
\nonumber
\\[3pt]
&& 
\phantom{1-\delta_{k,0} \ \ \ \ \ \ \ \  } \!\!{\geq}1 \ \ \ \ \ \ 0 \ \ \ \ \ \ \  0 \ \ \ \ \ \ \   0 \ \ \ \ \ \ {\geq}0 \ \ \ \ \,  {\geq}0   \ \ \ \ \  {\geq}0  
\nonumber
\\[3pt]
&& 
\phantom{1-\delta_{k,0} \ \ \ \ \ \ \ \  } 0 \ \ \ \ \ \  {\geq}1 \ \ \ \ \ \  0 \ \ \ \ \ \ \   0 \ \ \ \ \ \ {\geq}0 \ \ \ \ \,  {\geq}0   \ \ \ \ \  {\geq}0  
\nonumber
\\[3pt]
&& 
\phantom{1-\delta_{k,0} \ \ \ \ \ \ \ \  } 0 \ \ \ \ \ \ \ 0 \ \ \ \ \ \  {\geq}1 \ \ \ \ \ \   0 \ \ \ \ \ \ {\geq}0 \ \ \ \ \,  {\geq}0   \ \ \ \ \  {\geq}0  
\nonumber
\\[3pt]
&& 
\phantom{1-\delta_{k,0} \ \ \ \ \ \ \ \  } \!\!{\geq}1 \ \ \ \ \  {\geq}1 \ \ \ \ \ \,  0 \ \ \ \ \ \ \   0 \ \ \ \ \ \ {\geq}0 \ \ \ \ \,  {\geq}0   \ \ \ \ \  {\geq}0  
\nonumber
\\[3pt]
&& 
\phantom{1-\delta_{k,0} \ \ \ \ \ \ \ \  } \!\!{\geq}1 \ \ \ \ \ \, 0 \ \ \ \ \ \   {\geq}1 \ \ \ \ \ \    0 \ \ \ \ \ \ {\geq}0 \ \ \ \ \,  {\geq}0   \ \ \ \ \  {\geq}0  
\nonumber
\\[3pt]
&& 
\phantom{1-\delta_{k,0} \ \ \ \ \ \ \ \  } 0 \ \ \ \ \ \ {\geq}1 \ \ \ \ \,   {\geq}1 \ \ \ \ \ \     0 \ \ \ \ \ \ {\geq}0 \ \ \ \ \,  {\geq}0   \ \ \ \ \  {\geq}0  
\nonumber
\\[3pt]
&& 
\phantom{1-\delta_{k,0} \ \ \ \ \ \ \ \  } \!\!{\geq}1 \ \ \ \ \,  {\geq}1 \ \ \ \ \   {\geq}1 \ \ \ \ \ \,    0 \ \ \ \ \ \ {\geq}0 \ \ \ \ \,  {\geq}0   \ \ \ \ \  {\geq}0  
\nonumber \\
\label{b24}
\end{eqnarray}

($u$) For $u$ we have  
\begin{eqnarray}
&& 
\phantom{1-\delta_{k,0} \ \ \ \ \ \ \ \  } x \ \ \ \ \ \ \ y \ \ \ \ \ \ \ z \ \ \ \ \ \ \ r \ \ \ \ \ \ \ u \ \ \ \ \ \ \ v \ \ \ \ \ \ \ w     
\nonumber
\\[6pt]
&& 
1-\delta_{k,0} \ \ \ \ \ \ \ \   0 \ \ \ \ \ \ \  0 \ \ \ \ \ \ \ 0 \ \ \ \ \ \ \ \, 0 \ \ \ \ \ \ {\geq}0 \ \ \ \ \,  {\geq}0  \ \ \ \ \  {\geq}0   
\nonumber
\\[2pt]
&& 
\phantom{1-}\ \, \delta_{k,0} \ \ \ \ \ \ \ \  0 \ \ \ \ \ \ \ 0 \ \ \ \ \ \ \ 0\ \ \ \ \ \ \ \, 0 \ \ \ \ \ \ {\geq}0 \ \ \ \ \,  {\geq}1  \ \ \ \ \  {\geq}1 
\nonumber \\
\label{b26}
\end{eqnarray}
and similar arrays for $v$ and $w$.

Hence we obtain the seven equations for the probabilities $x$, $y$, $z$, $r$, $u$, $v$, and $w$,  
Eqs.~(\ref{b100})--(\ref{b140}).%


\medskip

\section{Equations for three layers with double overlaps}
\label{sa32}

For networks with double edges (and without triple edges) the list of cases where the node does belong to the giant component is shorter than the list of cases where it does belong  (out of $2^6=64$ combinations of $0$ or $\geq1$ for each type of edge, $45$ result in a surviving node and $19$ result in a pruned node).
Then, it is more convenient to work with $1-S$ instead of $S$:

($1-S$) For $1-S$ we have
\begin{eqnarray}
&& 
\  \ x \ \ \ \ \ \ \ x' \ \ \ \ \ \ x'' \ \ \ \ \ \, y \ \ \ \ \ \ \; \, y' \ \ \ \ \ \ y''
\nonumber
\\[6pt]
&& 
\ \; \, 0 \ \ \ \ \  {\geq}0\ \ \ \ \  {\geq}0\ \ \ \ \  \ \; \, 0 \ \ \ \ \  \ \; \, 0 \ \ \ \ \  {\geq}0 
\nonumber
\\[2pt]
&& 
{\geq} 0 \ \ \ \ \  \ \; \, 0\ \ \ \ \ {\geq}0\ \ \ \ \  \ \; \, 0 \ \ \ \ \  {\geq}1 \ \ \ \ \  \ \; \, 0 
\nonumber
\\[2pt]
&& 
{\geq} 1  \ \ \ \ \  \ \; \, 0\ \ \ \ \ {\geq}0\ \ \ \ \  \ \; \, 0 \ \ \ \ \  \ \; \, 0 \ \ \ \ \  \ \; \, 0 
\nonumber
\\[2pt]
&& 
{\geq}0 \ \ \ \ \  {\geq}0\ \ \ \ \  \ \; \, 0\ \ \ \ \  {\geq}1  \ \ \ \ \  \ \; \, 0 \ \ \ \ \  \ \; \,0 
\nonumber
\\[2pt]
&& 
{\geq}1 \ \ \ \ \  {\geq}1\ \ \ \ \  \ \; \, 0\ \ \ \ \  \ \; \, 0  \ \ \ \ \  \ \; \, 0 \ \ \ \ \  \ \; \,0 
\label{c10}
\end{eqnarray}
\begin{widetext}
Hence the expression for $1-S$ has the form:  
\begin{multline}
1 {-} S  = \!\!\!\!\!\!\!\!\!
\sum_{q,q',q'',k,k',k''} \!\!\!\!\!\!\!\!\! P(q,q',q'',k,k',k'')\Bigl\{
(1{-}x)^q (1{-}y)^{k} (1{-}y')^{k'}
+(1{-}x')^{q'} (1{-}y)^{k} [1{-}(1{-}y')^{k'}](1{-}y'')^{k''}
\\
+[1{-}(1{-}x)^q](1{-}x')^{q'} (1{-}y)^{k} (1{-}y')^{k'}(1{-}y'')^{k''}
+(1{-}x'')^{q''} [1{-}(1{-}y)^{k}] (1{-}y')^{k'}(1{-}y'')^{k''}
\\
+[1{-}(1{-}x)^q][1{-}(1{-}x')^{q'}](1{-}x'')^{q''} (1{-}y)^{k} (1{-}y')^{k'}(1{-}y'')^{k''},
\Bigr\}
\label{c20}
\end{multline}
which, when written more compactly, gives Eq. (\ref{c30}).

\end{widetext}

($1-x$) For $1-x$ we have
\begin{eqnarray}
&& 
\  \ x \ \ \ \ \ \ \ x' \ \ \ \ \ \ x'' \ \ \ \ \ \, y \ \ \ \ \ \ \; \, y' \ \ \ \ \ \ y''
\nonumber
\\[6pt]
&& 
{\geq} 0 \ \ \ \ \  \ \; \, 0\ \ \ \ \ {\geq}0\ \ \ \ \  \ \; \, 0 \ \ \ \ \  {\geq}0 \ \ \ \ \  \ \; \, 0 
\nonumber
\\[2pt]
&& 
{\geq} 0  \ \ \ \ \  {\geq}0 \ \ \ \ \  \ \; \, 0 \ \ \ \ \   {\geq}1 \ \ \ \ \  \ \; \, 0 \ \ \ \ \  \ \; \, 0 
\nonumber
\\[2pt]
&& 
{\geq} 0  \ \ \ \ \  {\geq}1  \ \ \ \ \  \ \; \, 0 \ \ \ \ \   \ \; \, 0  \ \ \ \ \  \ \; \, 0 \ \ \ \ \  \ \; \, 0 
\label{c40}
\end{eqnarray}
and two similar arrays for $x'$ and $x''$, leading to Eqs.~(\ref{c120})--(\ref{c140}).

($1-y$) For $1-y$ we have
\begin{eqnarray}
&& 
\  \ x \ \ \ \ \ \ \ x' \ \ \ \ \ \ x'' \ \ \ \ \ \, y \ \ \ \ \ \ \; \, y' \ \ \ \ \ \ y''
\nonumber
\\[6pt]
&& 
{\geq} 0 \ \ \ \ \  {\geq} 0  \ \ \ \ \ \ \; \, 0 \ \ \ \ \  {\geq}0  \ \ \ \ \  \ \; \, 0 \ \ \ \ \  \ \; \, 0 
\label{c50}
\end{eqnarray}
and two similar arrays for $y'$ and $y''$, leading to Eqs.~(\ref{c150})--(\ref{c170}).


\section{Critical behavior near the tricritical point in three layers with triple overlaps}\label{ap_C}

Beginning from the expansion of  $\tilde{F}(x)$ up to the cubic term, Eq. (\ref{c350}),
and writing $\langle q \rangle = \langle q \rangle_\text{T} + \delta$ and $\langle k \rangle_\text{T}+\tilde{\delta}$,
close to the tricritical point we can write
\begin{align}
\tilde{F}(0)&=0,
\nonumber
\\
\tilde{F}'(0) &\cong 1+A \delta + \tilde{A} \tilde{\delta}
\nonumber
\\
\frac{\tilde{F}''(0)}{2!}& \cong B \delta + \tilde{B} \tilde{\delta},
\nonumber
\\
\frac{\tilde{F}'''(0)}{3!} &\cong C,
\label{c360}
\end{align}
where
\begin{align}
A &= \frac{1+\sqrt{33}}{8}
,
\nonumber
\\
\tilde{A}  &= \frac{17+3\sqrt{33}}{4}
,
\nonumber
\\
B &= 0
, 
\nonumber
\\
\tilde{B} &= -3\frac{143+25\sqrt{33}}{16}
,
\nonumber
\\
C &= -\frac{6+\sqrt{33}}{6}
.
\label{c370}
\end{align}

Then, near the tricritical point, the self-consistency equation $\tilde{F}(x)=x$ can be expressed as
\begin{equation}
(A\delta + \tilde{A}\tilde{ \delta})x + \tilde{B} \tilde{\delta} x^2 + C x^3 = 0,
\label{c380}
\end{equation}
which, besides the solution $x=0$, has two other solutions
\begin{equation}
x = {-} \frac{  \tilde{B}  \tilde{\delta} \pm \sqrt{ { \tilde{B}}^2 { \tilde{\delta}}^2 - 4 C(A\delta +  \tilde{A}  \tilde{\delta})  }}{2 C}
.
\label{c390}
\end{equation}

The discontinuous transition occurs when the two non-trivial solutions of Eq.~(\ref{c390}) are equal, i.e., when the argument of the square root in Eq.~(\ref{c390}) is zero, which leads to
\begin{align}
 \tilde{\delta}_{c} &\cong 2\frac{ C \tilde{A}   + \sqrt{C^2 {\tilde{A}  }^2+ C A {\tilde{B}  }^2  \delta } }{ {\tilde{B}  }^2}
 \nonumber \\
&\cong-\frac{ A}{ \tilde{A}   } \delta =  -\frac{41-7\sqrt{33}}{8} \delta
.
\label{c400}
\end{align}
Notice that the slope of the line of the discontinuity at the tricritical point is equal to the slope of the line of continuous transitions at the same point, i.e., $\partial_{\langle q \rangle} \langle k \rangle_c^\text{disc} (\langle q \rangle_\text{T}) = \partial_{\langle q \rangle} \langle k \rangle_c^\text{cont} (\langle q \rangle_\text{T})  = -{[41-7\sqrt{33}]}/{8}$.

The sizes of the jumps in $x$ and $y$ are
\begin{equation}
x_c \cong  \frac{  A   \tilde{B}}{2  \tilde{A} C} \delta   = 3\frac{7\sqrt{33}-33}{16} \delta  
\label{c410}
\end{equation}
and
\begin{equation}
y_c \cong   \frac{  A   \tilde{B}}{2 \langle k \rangle_\text{T}  \tilde{A}C} \delta   = \frac{3 \sqrt{33}}{4} \delta  .
\label{c420}
\end{equation}

As for the jump in $S$, the expansion of Eq.~(\ref{c260}) gives
\begin{equation}
S =  3 \langle k \rangle^2 y^2 + 3 \langle q \rangle \langle k \rangle x y + ... ,
\label{c430}
\end{equation}
then
\begin{equation}
S_c \cong
\frac{891}{128}(5\sqrt{33}-27){\delta }^2
\label{c440}
\end{equation}

Let us check the critical exponents of the hybrid transition.
Letting $ \epsilon  =  \langle q \rangle - \langle q \rangle_c$ and $ \tilde{\epsilon}= \langle k \rangle - \langle k \rangle_c$ be deviations from a point $ (\langle q \rangle_c, \langle k \rangle_c)$ in the critical line. 
For $ (\langle q \rangle_c, \langle k \rangle_c)$ near the tricritical point, and for small $ \epsilon $ and $ \epsilon_{\langle k \rangle}$, from Eq.~(\ref{c390}) we get
\begin{eqnarray}
x &=& x_c + \sqrt{ {-} \frac{ \tilde{A} \tilde{\epsilon}{+}  A  \epsilon } {C} } + ...
.
\nonumber
\\
&\cong & x_c + \sqrt{   \frac{3{+}\sqrt{33}}{2}  \tilde{\epsilon}{+} \frac{5 \sqrt{33} {-} 27}{4} \epsilon  }
\label{c450}
\end{eqnarray}
For $y$ we simply divide $x$ by $\langle k \rangle_\text{T}$,
\begin{equation}
y \cong y_c + \frac{7{+}\sqrt{33}}{4}  \sqrt{   \frac{3{+}\sqrt{33}}{2}  \tilde{\epsilon}{+} \frac{5 \sqrt{33} {-} 27}{4} \epsilon  }
.
\label{c460}
\end{equation}
Near the tricritical point the expansion of $S$ is
\begin{eqnarray}
S &\cong& S_c + 27 \frac{11{-}\sqrt{33}}{8}  \delta  \sqrt{   \frac{3{+}\sqrt{33}}{2}  \tilde{\epsilon}{+} \frac{5 \sqrt{33} {-} 27}{4} \epsilon  }
\nonumber
\\
&+&  3 \frac{3+\sqrt{33}}{4} \left( \frac{3{+}\sqrt{33}}{2}  \tilde{\epsilon}{+} \frac{5 \sqrt{33} {-} 27}{4} \epsilon \right)
.
\label{c470}
\end{eqnarray}

Notice that in Eqs.~(\ref{c450})--(\ref{c470}), that is, Eq.~(\ref{c470a}), the argument of the square root is positive as long as we remain above the transition line.
Furthermore, the amplitudes of the square-root singular terms in Eqs.~(\ref{c450}) and (\ref{c460}), for $x$ and $y$, remain finite at the tricritical point, but the one of Eq.~(\ref{c470}), for $S$, is linear in $ \delta  $. 
So, the region where the square root of $S$ dominates over the linear contributions vanishes approaching the tricritcal point, and, at the tricritical point, the singularity of $S$ has exponent $1$.


Near the point ($\langle q \rangle_c, \langle k \rangle_c ) $ on the line of continuous transitions, Eq.~(\ref{c291}), we have 
\begin{eqnarray}
&&
\tilde{F}(0)=0,
\nonumber
\\
&&
\tilde{F}'(0) \cong 1 + \frac{1}{\langle q \rangle_c}  \epsilon   +  \frac{(2+\langle q \rangle_c)^2}{\langle q \rangle_c}   \tilde{\epsilon}  ,
\nonumber
\\
&&
\tilde{F}''(0) \cong  -\frac{(1{+}\langle q \rangle_c)(4{-}\langle q \rangle_c{-} 2 {\langle q \rangle_c}^2)}{\langle q \rangle_c}  
,
\label{c480}
\end{eqnarray}
where $\epsilon  = {\langle q \rangle} - {\langle q \rangle}_c $ and $ \tilde{\epsilon}= {\langle k \rangle} - {\langle k \rangle}_c $, similarly to before.
Solving $x=\tilde{F}(0) + \tilde{F}'(0) x+ \tilde{F}''(0) x^2/2 $ gives
\begin{equation}
x =  
 \frac{ 2 }{ (1 {+} {\langle q \rangle_c} ) (4 {-} \langle q \rangle_c {-} 2 {\langle q \rangle_c}^2 ) }  [ \epsilon  + (2 {+} \langle q \rangle_c)^2   \tilde{\epsilon}]
,
\label{c490}
\end{equation}
\begin{equation}
y =  
 \frac{ 4 + 2{\langle q \rangle_c}   }{ (1 {+} {\langle q \rangle_c} ) (4 {-} \langle q \rangle_c {-} 2 {\langle q \rangle_c}^2 ) }  [ \epsilon  + (2 {+} \langle q \rangle_c)^2   \tilde{\epsilon}]
 ,
\label{c500}
\end{equation}
and 
\begin{equation}
S =  
 \frac{ 12(1 + {\langle q \rangle_c}) }{ {(1 {+} {\langle q \rangle_c} )}^2 {(4 {-} \langle q \rangle_c {-} 2 {\langle q \rangle_c}^2 )}^2 }  {[ \epsilon  + (2 {+} \langle q \rangle_c)^2   \tilde{\epsilon}]}^2
 ,
\label{c510}
\end{equation}
that is, Eq.~(\ref{c510a}).

\bibliographystyle{elsarticle-num} 
\bibliography{weak_overlaps_refs}

\begin{thebibliography}{10}
\expandafter\ifx\csname url\endcsname\relax
  \def\url#1{\texttt{#1}}\fi
\expandafter\ifx\csname urlprefix\endcsname\relax\def\urlprefix{URL }\fi
\expandafter\ifx\csname href\endcsname\relax
  \def\href#1#2{#2} \def\path#1{#1}\fi

\bibitem{newman2010networks}
M.~E.~J. Newman, Networks: An Introduction, Oxford University Press, Oxford,
  2010.

\bibitem{dorogovtsev2008critical}
S.~N. Dorogovtsev, A.~V. Goltsev, J.~F.~F. Mendes, Critical phenomena in
  complex networks, Rev. Mod. Phys. 80 (2008) 1275.

\bibitem{dorogovtsev2022the}
S.~N. Dorogovtsev, J.~F.~F. Mendes, The Nature of Complex Networks, Oxford
  University Press, Oxford, 2022.

\bibitem{buldyrev2010catastrophic}
S.~V. Buldyrev, R.~Parshani, R.~Paul, H.~E. Stanley, S.~Havlin, Catastrophic
  cascade of failures in interdependent networks, Nature 464 (2010) 1025.

\bibitem{baxter2012avalanche}
G.~J. Baxter, S.~N. Dorogovtsev, A.~V. Goltsev, J.~F.~F. Mendes, Avalanche
  collapse of interdependent networks, Phys. Rev. Lett. 109 (2012) 248701.

\bibitem{baxter2014weak}
G.~J. Baxter, S.~N. Dorogovtsev, J.~F.~F. Mendes, D.~Cellai, Weak percolation
  on multiplex networks, Phys. Rev. E 89~(4) (2014) 042801.

\bibitem{bianconi2018multilayer}
G.~Bianconi, Multilayer Networks: Structure and Function, Oxford University
  Press, Oxford, 2018.

\bibitem{baxter2020exotic}
G.~J. Baxter, R.~A. da~Costa, S.~N. Dorogovtsev, J.~F.~F. Mendes, Exotic
  critical behavior of weak multiplex percolation, Phys. Rev. E 102~(3) (2020)
  032301.

\bibitem{baxter2021weak}
G.~J. Baxter, R.~A. da~Costa, S.~N. Dorogovtsev, J.~F.~F. Mendes, Weak
  Multiplex Percolation, Elements in Structure and Dynamics of Complex
  Networks, Cambridge University Press, Cambridge, 2021.

\bibitem{baxter2015critical}
G.~J. Baxter, S.~N. Dorogovtsev, K.-E. Lee, J.~F.~F. Mendes, A.~V. Goltsev,
  Critical dynamics of the $k$-core pruning process, Phys. Rev. X 5~(3) (2015)
  031017.

\bibitem{azimi2014k}
N.~Azimi-Tafreshi, J.~G{\'o}mez-Garde{\~n}es, S.~N. Dorogovtsev, ${k}$-core
  percolation on multiplex networks, Phys. Rev. E 90~(3) (2014) 032816.

\bibitem{di2017cascading}
M.~A. Di~Muro, L.~D. Valdez, H.~H. Arag{\~a}o~R{\^e}go, S.~V. Buldyrev, H.~E.
  Stanley, L.~A. Braunstein, Cascading failures in interdependent networks with
  multiple supply-demand links and functionality thresholds, Sci. Reports 7
  (2017) 1.

\bibitem{shao2011cascade}
J.~Shao, S.~V. Buldyrev, S.~Havlin, H.~E. Stanley, Cascade of failures in
  coupled network systems with multiple support-dependence relations, Phys.
  Rev. E 83~(3) (2011) 036116.

\bibitem{son2012percolation}
S.-W. Son, G.~Bizhani, C.~Christensen, P.~Grassberger, M.~Paczuski, Percolation
  theory on interdependent networks based on epidemic spreading, EPL 97~(1)
  (2012) 16006.

\bibitem{hu2013percolation}
Y.~Hu, D.~Zhou, R.~Zhang, Z.~Han, C.~Rozenblat, S.~Havlin, Percolation of
  interdependent networks with intersimilarity, Phys. Rev. E 88~(5) (2013)
  052805.

\bibitem{min2015link}
B.~Min, S.~Lee, K.-M. Lee, K.-I. Goh, Link overlap, viability, and mutual
  percolation in multiplex networks, Chaos, Solitons \& Fractals 72 (2015) 49.

\bibitem{baxter2016correlated}
G.~J. Baxter, G.~Bianconi, R.~A. da~Costa, S.~N. Dorogovtsev, J.~F.~F. Mendes,
  Correlated edge overlaps in multiplex networks, Phys. Rev. E 94~(1) (2016)
  012303.

\bibitem{cellai2016message}
D.~Cellai, S.~N. Dorogovtsev, G.~Bianconi, Message passing theory for
  percolation models on multiplex networks with link overlap, Phys. Rev. E
  94~(3) (2016) 032301.

\bibitem{cellai2011tricritical}
D.~Cellai, K.~A. Lawlor, A.and~Dawson, J.~P. Gleeson, Tricritical point in
  heterogeneous $k$-core percolation, Phys. Rev. Lett. 107~(17) (2011) 175703.

\bibitem{baxter2011heterogeneous}
G.~J. Baxter, S.~N. Dorogovtsev, A.~V. Goltsev, J.~F.~F. Mendes, Heterogeneous
  $k$-core versus bootstrap percolation on complex networks, Phys. Rev. E
  83~(5) (2011) 051134.

\end{thebibliography}

\end{document}